\documentstyle[a4p,12pt,epsfig,feynmf,cite]{article}

\newcommand{\sqee}{\sqrt{s_{\rm ee}}}
\newcommand{\epem}{$\mbox{e}^+\mbox{e}^-$}
\def\gaga{\gamma\gamma}

\def\sqee{\sqrt{s}_{\rm ee}}

\def\pt{p_{\rm T}}

\def\Wvis{W_{\rm vis}}

\def\qmax{Q^2_{\rm max}}
\def\qmin{Q^2_{\rm min}}

\def\ccbar{\mbox{c}\overline{\mbox{c}}}

\def\pz{\phantom{0}}

\def\BR{\mbox{BR}}
\def\DSTPM{{\rm D}^{\ast\pm}}
\def\DSTP{{\rm D}^{\ast+}}
\def\DSTM{{\rm D}^{\ast-}}
\def\DST{{\rm D}^{\ast}}
\def\D0{{\rm D}^0}
\def\DELTACAND{\Delta M \equiv M_{\DST}^{\rm cand}-M_{\D0}^{\rm cand}}
\def\DELTAM{\Delta M \equiv M_{\DST}-M_{\D0}}
\def\ptdst{p_{\rm T}^{\DST}}
\def\etadst{\eta^{\DST}}
\def\dspt{{\rm d}\sigma/{\rm d}p_{\rm T}^{\DST}}
\def\dseta{{\rm d}\sigma/{\rm d}|\etadst|}
\def\ndst{N_{\DST}^{\rm rec}}

\def\xgmin{x^{\rm min}_{\gamma}}
\def\K{\mbox{K}}
\def\gaga{\gamma\gamma}
\def\sigcc{\sigma({\rm e}^+{\rm e}^-\to{\rm e}^+{\rm e}^-{\rm c}\bar{{\rm c}})}
\def\sigdst{\sigma({\rm e}^+{\rm e}^-\to{\rm e}^+{\rm e}^- \DST X)}
\def\CF2{F_{\rm 2,c}^{\gamma}}
\def\XQCF2{F_{\rm 2,c}^{\gamma}(x,\langle Q^2\rangle)}
\def\F2{F_2^{\gamma}}
\def\ntag{29.8 \pm 5.9~({\rm stat})}
\def\sigtota{842\pm 97~({\rm stat})\pm  75~({\rm sys})\pm 
196~({\rm extrapolation})~{\rm pb}}
\def\sigtot{842\pm 97~({\rm stat})\pm  75~({\rm sys})\pm 
196~({\rm extr})~{\rm pb}}
\def\sigdir{351\pm 40~({\rm stat})\pm  79~({\rm sys})\pm
 66~({\rm extr})~{\rm pb}}
\def\sigres{491\pm 56~({\rm stat})\pm 111~({\rm sys})\pm
130~({\rm extr})~{\rm pb}}
\def\xt{x_{\rm T}^{\DST}}
\begin{document}
\begin{titlepage}
\begin{center}{\large   EUROPEAN LABORATORY FOR PARTICLE PHYSICS
}\end{center}\bigskip
\begin{flushright}
       CERN-EP/99-157   \\ 9 November 1999
\end{flushright}
\bigskip\bigskip\bigskip\bigskip\bigskip
\begin{center}{\huge\bf \boldmath  
Inclusive Production of D$^{\ast\bf\pm}$ Mesons in 
Photon-Photon Collisions at $\sqrt{s}_{\rm\bf ee}=183$
and $189$ GeV and \\ a First Measurement of $\CF2$ \unboldmath
}\end{center}\bigskip\bigskip
\begin{center}{\LARGE The OPAL Collaboration
}\end{center}\bigskip\bigskip
\bigskip\begin{center}{\large  Abstract}\end{center}
The inclusive production of $\DSTPM$ mesons in photon-photon collisions
has been measured using the OPAL detector at LEP at \epem~centre-of-mass
energies $\sqee$ of $183$ and 189~GeV. 
The $\DSTP$ mesons are reconstructed in their
decay to ${\rm D}^0\pi^+$ with the ${\rm D}^0$ observed in the
two decay modes ${\rm K}^-\pi^+$ and ${\rm K}^-\pi^+\pi^-\pi^+$. 
After background subtraction,
\mbox{$100.4\pm12.6~(\rm stat)$} $\DSTPM$ mesons have 
been selected in events without observed scattered beam electron
(``anti-tagged'') 
and $\ntag~\DSTPM$ mesons in events
where one beam electron is scattered into the detector (``single-tagged''). 
Direct and single-resolved events are studied separately.  
Differential cross-sections $\dspt$ and $\dseta$
as functions of the $\DSTPM$ transverse momentum $\ptdst$ 
and pseudorapidity $\etadst$ are presented in 
the kinematic region $2~{\rm GeV}<\ptdst<12~{\rm GeV}$ and $|\etadst|<1.5$. 
They are compared 
to next-to-leading order (NLO) perturbative QCD calculations.
The total cross-section for the process \epem $\to$ \epem $\ccbar$
where the charm quarks are produced 
in the collision of two quasi-real photons
is measured to be $\sigcc = \sigtota$.
A first measurement of the charm structure function $\CF2$ of the photon 
is performed in the kinematic range $0.0014<x<0.87$ 
and $5~{\rm GeV}^2<Q^2<100~{\rm GeV}^2$, 
and the result is compared to a NLO perturbative QCD calculation. 
\bigskip\bigskip\bigskip\bigskip
\bigskip\bigskip
\begin{center}{\large
(To be submitted to Eur.~Phys.~J.~C)
}\end{center}
\end{titlepage}
\begin{center}{\Large        The OPAL Collaboration
}\end{center}\bigskip
\begin{center}{
G.\thinspace Abbiendi$^{  2}$,
K.\thinspace Ackerstaff$^{  8}$,
P.F.\thinspace Akesson$^{  3}$,
G.\thinspace Alexander$^{ 23}$,
J.\thinspace Allison$^{ 16}$,
K.J.\thinspace Anderson$^{  9}$,
S.\thinspace Arcelli$^{ 17}$,
S.\thinspace Asai$^{ 24}$,
S.F.\thinspace Ashby$^{  1}$,
D.\thinspace Axen$^{ 29}$
% ...
}\end{center}\bigskip
\bigskip
$^{  1}$School of Physics and Astronomy, University of Birmingham,
Birmingham B15 2TT, UK
\newline
$^{  2}$Dipartimento di Fisica dell' Universit\`a di Bologna and INFN,
I-40126 Bologna, Italy
\newline
$^{  3}$Physikalisches Institut, Universit\"at Bonn,
D-53115 Bonn, Germany
\newline
$^{  4}$Department of Physics, University of California,
Riverside CA 92521, USA
\newline
% ...
$^{  6}$Ottawa-Carleton Institute for Physics,
Department of Physics, Carleton University,
Ottawa, Ontario K1S 5B6, Canada
\newline
% ...
$^{  8}$CERN, European Organisation for Particle Physics,
CH-1211 Geneva 23, Switzerland
\newline
$^{  9}$Enrico Fermi Institute and Department of Physics,
University of Chicago, Chicago IL 60637, USA
\newline
% ...
$^{ 12}$Indiana University, Department of Physics,
Swain Hall West 117, Bloomington IN 47405, USA
\newline
% ...
$^{ 16}$Department of Physics, Schuster Laboratory, The University,
Manchester M13 9PL, UK
\newline
$^{ 17}$Department of Physics, University of Maryland,
College Park, MD 20742, USA
\newline
% ...
$^{ 23}$Department of Physics and Astronomy, Tel Aviv University,
Tel Aviv 69978, Israel
\newline
$^{ 24}$International Centre for Elementary Particle Physics and
Department of Physics, University of Tokyo, Tokyo 113-0033, and
Kobe University, Kobe 657-8501, Japan
\newline
% ...
$^{ 29}$University of British Columbia, Department of Physics,
Vancouver BC V6T 1Z1, Canada
\newline
% ...

\section{Introduction}

Charged $\DST$ mesons\footnote{Throughout
this paper $\DST$ refers to $\DSTP$ as well as to $\DSTM$.
Charge conjugated modes are always implied.}
provide a clean tag to study open charm production in
photon-photon collisions.
The inclusive cross-section for the production
of $\DST$ mesons can be calculated in
perturbative QCD (pQCD). Since the process is characterised by
two distinct scales, the mass $m_{\rm c}$
and the transverse momentum $p_{\rm T}$ of the charm quarks,
two different approaches exist for the
next-to-leading order (NLO) pQCD calculations.
In the ``massless'' scheme,
charm is treated as an active flavour in the photon. This approach is
expected to be valid for $p_{\rm T} \gg m_{\rm c}$.
In the ``massive'' scheme, matrix elements for massive charm quarks are used
and no charm content is assigned to the parton distributions of
the photon. One expects this ansatz to be valid
at $p_{\rm T} \approx m_{\rm c}$~\cite{bib-mass}.

A third scale is the 
four-momentum squared, $Q_{i=1,2}^2$, of the interacting virtual photons. 
In this paper, two kinematic cases are studied, depending on $Q_{i}^2$. 
For the largest part of the cross-section, both
exchanged photons are quasi-real ($Q_1^2,Q_2^2\approx0$) and
the beam electrons are scattered at very small angles. 
These events are selected by rejecting events with 
a scattered electron in the detector. 
Events of this type are called anti-tagged.
If a photon is highly virtual ($Q_i^2 \gg 0$), 
the corresponding beam electron is usually scattered
into the acceptance of the detector. Events with one detected 
scattered electron are called single-tagged.

In direct events, the two photons couple directly to the $\ccbar$ pair.
In resolved events, one photon (``single-resolved'') or
both photons (``double-resolved'') fluctuate into a hadronic state and
a gluon or a quark of the hadronic
fluctuation of the photon takes part in the hard interaction.
For anti-tagged events at \epem~centre-of-mass energies $\sqee$ 
around 189 GeV, the production of $\DST$ 
mesons in photon-photon collisions in leading order (LO)
QCD proceeds mainly via direct 
($\gaga\to\ccbar$) and single-resolved (${\rm g}\gamma\to\ccbar$) 
photon-photon processes~\cite{bib-DKZZ,bib-Frixione}, whereas 
the contribution from double-resolved events
(${\rm gg}\to\ccbar$) is expected to be small. 
The measurement of the open charm cross-section is
therefore expected to be sensitive to the gluon content of the photon
through the photon-gluon fusion process ${\rm g}\gamma \to \ccbar$.
The production of $\DST$ mesons from open bottom production 
in photon-photon events is expected to be suppressed by more
than two orders of magnitude~\cite{bib-DKZZ}. 
Bottom production is suppressed due to the smaller electric charge and
the larger mass of the b quarks.

The single-tagged process can be regarded as deep inelastic electron-photon
scattering. 
In this configuration, the electron radiating the highly virtual photon, 
$\gamma^*$, probes the structure of the quasi-real photon, $\gamma$, 
radiated from the second beam electron, and allows the determination
of the photon structure function $F^{\gamma}_2(x,Q^2)$.
Here $x$ is the Bjorken scaling variable and $Q^2$ the virtuality
of the highly virtual photon.
Due to the large mass of the charm quark, the charm structure function
$\CF2$ of the photon can be calculated in pQCD to 
NLO~\cite{bib-Laenen}.
In QCD, $\CF2$ receives contributions from the point-like and the 
hadron-like structure of the quasi-real photon. 
The two contributions are expected to be well separated in $x$, with
the point-like contribution dominating at high $x$ and the 
hadron-like component sizeable only for $x<0.1$.

The production of $\DST$ mesons in photon-photon events has been measured
previously by JADE~\cite{bib-JADE}, TASSO~\cite{bib-TASSO}, 
TPC/2$\gamma$~\cite{bib-TPC},
TOPAZ~\cite{bib-TOPAZ}, AMY~\cite{bib-AMY}, ALEPH~\cite{bib-ALEPH} and 
L3~\cite{bib-L3DST} at $\sqee$ ranging from 29~GeV up to 189~GeV.
The charm structure function $\CF2$ of the photon has never
been measured before. 
The analysis presented here uses the data taken with the OPAL 
detector in 1997 at
$\sqee=183~{\rm GeV}$ and in 1998 at $\sqee=189~{\rm GeV}$
with integrated luminosities $\cal{L}$ of 55 and 165~pb$^{-1}$,
respectively.

\section{The OPAL detector}

A detailed description of the OPAL detector
can be found in Ref.~\cite{bib-opaltechnicpapers}, and
therefore only a brief account of the main features relevant
to the present analysis will be given here.

The central tracking system is located inside a solenoidal magnet which
provides a uniform axial magnetic field of 0.435~T along the beam
axis\footnote{In the OPAL coordinate system the $x$ axis points
  towards the centre of the LEP ring, the $y$ axis points upwards and
  the $z$ axis points in the direction of the electron beam.  The
  polar angle $\theta$ is defined with respect to the $z$ axis.}.
The magnet is surrounded by a lead-glass electromagnetic
calorimeter (ECAL) and a hadronic sampling calorimeter (HCAL).
Outside the HCAL, the detector is surrounded by muon
chambers. There are similar layers of detectors in the
endcaps. The region around the beam pipe on both sides
of the detector is covered by the forward calorimeters and the
silicon-tungsten luminometers.

Starting with the innermost components, the
tracking system consists of a high precision silicon
microvertex detector (SI), a precision vertex
drift chamber (CV), a large volume jet chamber (CJ) with 159 layers of axial
anode wires and a set of $z$ chambers measuring the track coordinates
along the beam direction.
The transverse momenta $\pt$ of tracks are measured with a precision of
$\sigma_{\pt}/\pt=\sqrt{0.02^2+(0.0015\cdot \pt)^2}$ 
($\pt$ in GeV)\footnote{Throughout this paper we use the convention $c=1$.}
in the central region $|\cos\theta|<0.73$. In this paper, transverse
is always defined with respect to the $z$ direction of the detector.
The jet chamber also provides energy loss
measurements which are used for particle identification.

The ECAL completely covers the azimuthal range for polar angles
satisfying $|\cos\theta|<0.98$. The barrel section, which covers
the polar angle range $|\cos\theta|<0.82$, consists of a cylindrical array
of 9440 lead-glass blocks with a depth of
$24.6$ radiation lengths. The endcap sections consist of 1132 
lead-glass blocks with a depth of more than $22$ radiation lengths, 
covering the polar angle between $0.81 < |\cos\theta| < 0.98 $.

The forward calorimeters (FD) at each end of the OPAL detector
consist of cylindrical lead-scintillator calorimeters with a depth of
24 radiation lengths divided azimuthally into 16 segments.
The electromagnetic energy resolution is about
$18\%/\sqrt{E}$~($E$ in GeV).
The acceptance of the forward calorimeters covers the angular range
from 47 to 140~mrad from the beam direction.
Three planes of proportional tube chambers at 4 radiation lengths
depth in the calorimeter measure the directions of electron showers
with a precision of approximately 1~mrad.

The silicon tungsten detectors (SW)~\cite{bib-opalsiw} at each end of the OPAL
detector cover an angular region between 33 and 59 mrad
in front of the forward calorimeters.
Each calorimeter consists of 19 layers of silicon detectors and 18
layers of tungsten, corresponding to a total of 22 radiation
lengths. Each silicon layer consists of 16 wedge-shaped 
silicon detectors. The electromagnetic energy resolution is about
$25\%/\sqrt{E}$ ($E$ in GeV). The radial position of electron
showers in the SW calorimeter can be determined with a
typical resolution of 0.06 mrad in the polar angle $\theta$.

\section{Process kinematics} 

The kinematic properties of the two interacting photons are described by their
negative squared four-mo\-mentum transfers, $Q^2_{i=1,2}$, which
are related to the scattering angles $\theta'_i$ 
relative to the beam direction of the corresponding electrons by
\begin{equation}
Q^2_i  =-(k_i-k'_i)^2\approx 2E_i E'_i(1-\cos\theta'_i),
\label{eq-q2}
\end{equation}
neglecting the mass $m_{\rm e}$ of the electron.
The quantities $k_i$ and $k'_i$ are the four-momenta of the beam
and scattered electrons, 
and $E_i$ and $E'_i$ are their respective energies. 
The flux of transversely polarized quasi-real photons  with 
an energy fraction $z$ of the beam 
energy and a negative squared four-momentum, denoted with $Q^2_i$, may 
be obtained by the Equivalent Photon Approximation (EPA)~\cite{bib-EPA}:
\begin{equation}
\frac{{\rm d}^2 N_{\gamma}}{{\rm d}z {\rm d} Q^2_i} = 
\frac{\alpha}{2\pi}\left(\frac{1+(1-z)^2}{z~Q^2_i}
-\frac{2m^2_{\rm e} z}{Q^4_i}\right),
\end{equation}
where $\alpha$ is the fine structure constant.
The minimum kinematically allowed squared four-momentum
transfer, $\qmin$, is determined by
the electron mass,
\begin{equation}
\qmin=\frac{m_{\rm e}^2z^2}{1-z}.
\end{equation}
The maximum squared four-momentum transfer, $\qmax$, is  
given by the experimental anti-tagging condition
according to Eq.~\ref{eq-q2}.

If one of the photons is highly virtual ($Q^2_1 \equiv Q^2 =-q^2
\gg Q^2_2 \equiv P^2 =-p^2\approx 0$),
the cross-section of the process $\rm e^+e^- \to e^+e^- {\rm c}\bar{\rm c}$ can be written
as a product of the deep inelastic electron-photon scattering
cross-section and the flux of quasi-real photons, 
\begin{equation}
\frac{{\rm d}^4\sigma_{\rm e^+e^- \to e^+e^- {\rm c}\bar{\rm c}}}{{\rm d}x{\rm d}Q^2 {\rm d}z {\rm d}P^2}=
     \frac{{\rm d}^2\sigma_{\rm e\gamma \to e {\rm c}\bar{\rm c}}}{{\rm d}x{\rm d}Q^2} \cdot
     \frac{{\rm d}^2 N_{\gamma}}{{\rm d}z {\rm d} P^2}.  
\end{equation}
The differential deep inelastic electron-photon scattering cross-section may be 
parametrised in terms of structure functions as~\cite{bib-Berger}
\begin{equation}
\frac{{\rm d}^2\sigma_{\rm e\gamma \to e {\rm c}\bar{\rm c}}}{{\rm d}x{\rm d}Q^2}=
     \frac{2\pi\alpha^2}{xQ^4} \cdot
     \left[(1+(1-y)^2) \CF2(x,Q^2 ) -y^2 F_{\rm L, c}^{\gamma}(x,Q^2)\right],
\label{eq-sig}
\end{equation}
where $x=Q^2/2pq$ and the 
inelasticity $y=pq/p k_1$ are the usual dimensionless
deep inelastic scattering variables. 
In the kinematic regime studied in this paper,
$y^2 \ll 1$. The contribution proportional to $F_{\rm L, c}^{\gamma}(x,Q^2)$
in Eq.~\ref{eq-sig} is therefore neglected.
The variable $x$ is experimentally accessible via the relation 
\begin{eqnarray}
x=\frac{Q^2}{Q^2+W^2+P^2}\approx\frac{Q^2}{Q^2+W^2},
\label{eq-x}
\end{eqnarray}
where $W^2$ is the invariant mass squared
of the photon-photon system. 
The event selection ensures that the virtuality $P^2$ 
is usually very small compared to $Q^2$, 
so $P^2$ is neglected for the determination of $x$ from Eq.~\ref{eq-x}. 
By measuring the deep inelastic electron-photon scattering
cross-section for $\rm e^+e^- \to e^+e^- {\rm c}\bar{\rm c}$
as a function of $x$ and $Q^2$,
the charm structure function $\CF2(x,Q^2 )$ of the photon can
be determined.

\section{Monte Carlo simulation}

For real photons the PYTHIA 6.121~\cite{bib-pythia} photon-photon Monte Carlo
program, based on LO pQCD calculations, is used to 
simulate the process 
\epem $\to$ \epem $\gamma\gamma\to$ \epem $\ccbar\to$ \epem $\DST X$  
($X$ is a hadronic system).
Two distinct samples, one for the
direct process, $\gaga\to\ccbar$, and one for 
the single-resolved process, ${\rm g}\gamma\to\ccbar$, were generated using 
matrix elements for massive charm quarks. The different \epem~centre-of-mass
energies, $\sqee = 183$ and $189$~GeV, are taken into account by generating 
events at both energies according  
to the ratio of the corresponding integrated luminosities.
In case of the single-resolved process, 
the SaS-1D parametrisation~\cite{bib-sas} 
is used 
for the parton distributions of the photon. The fragmentation
of the charm quarks is modelled using the Peterson
fragmentation function~\cite{bib-Peterson} 
with the PYTHIA default parameter $\epsilon_{\rm c}=0.031$, and the charm
mass is set to $m_{\rm c}=1.6~{\rm GeV}$. 
The resulting average scaled energy of the generated 
$\DST$ mesons is
$\langle x_{\DST}\rangle= 2\cdot \langle E_{\DST}/W \rangle=0.84$, 
where $E_{\DST}$ is the 
$\DST$ energy in the photon-photon centre-of-mass system.
Final state QCD radiation off the primary charm quarks is taken 
into account using the leading-log approximation.  
A sample of double-resolved quasi-real photon-photon
events (${\rm gg}\to\ccbar$) was also generated with the PYTHIA 6.121 Monte
Carlo generator.

 The LO Monte Carlo generators HERWIG 5.9~\cite{bib-HERWIG} and
 Vermaseren~\cite{bib-Vermaseren} are used to model the 
 $\DST$ production in deep inelastic electron-photon scattering,
 \epem $\to$ \epem $\gamma^*\gamma\to$ \epem $\ccbar\to$ \epem $\DST X$. 
 For both Monte Carlo generators, the charm quark mass is chosen to be 
 $m_{\rm c}=1.5$~GeV.
 In HERWIG, the cross-section is evaluated for massless charm quarks.
 The charm production is modelled using matrix elements for massless
 charm quarks, together with the GRV parametrisation~\cite{bib-grv} 
 for the parton distributions of the photon, again for massless charm quarks.
 The effect of the charm quark mass is accounted for rather crudely by not
 simulating events with $W<2m_{\rm c}$, giving an unphysically sharp 
 step in the cross-section at this threshold.
 Due to the massless approach used in HERWIG and the crude treatment at 
 threshold, the
 predicted charm production cross-section is likely to be too large.
 Nevertheless, the final state kinematics are treated correctly, so
 HERWIG can be used for the determination of selection efficiencies.
 The fragmentation of quarks into hadrons is modelled via the cluster
 fragmentation model yielding an average scaled energy of the generated
 $\DST$ mesons of $\langle x_{\DST}\rangle=0.64$.
 The Vermaseren generator is based on the Quark Parton Model (QPM)
 and consequently does not take into account the hadron-like component of the 
 photon structure.
 It models the complete dependence of the cross-section on the different
 photon helicities. 
 The fragmentation into hadrons is handled via JETSET~7.4~\cite{bib-pythia},
 where the same fragmentation model is
 used as for the PYTHIA Monte Carlo described above.
 The average scaled energy of the generated $\DST$ mesons
 is $\langle x_{\DST}\rangle=0.82$, close to the PYTHIA value.

The \epem~annihilation background with $\DST$ mesons in the final state
has been simulated with the PYTHIA 5.7 Monte Carlo model. 
The Monte Carlo generator GRC4F~\cite{bib-GRC4F}
was used to simulate four-fermion processes 
that are background to the photon-photon sample.
All Monte Carlo samples were generated
with full simulation of the OPAL detector \cite{bib-gopal}.
They are analysed using the same reconstruction algorithms as
applied to the data.

\section{Event selection and \boldmath $\DST$ \unboldmath reconstruction}

In this and in the following three sections, only anti-tagged 
photon-photon scattering events are studied; the analysis of
tagged electron-photon scattering events is described in section 9.
Because the difference of the \epem~centre-of-mass energies of 
$\sqee=183$ and 189~GeV is small the 
data samples recorded at both energies are combined.
Anti-tagged photon-photon events are selected using the following set of cuts:
\begin{itemize} 
\item At least three tracks must have been found in the tracking chambers 
      (SI, CV and CJ). A track is
      required to have a minimum transverse momentum 
      of 120~MeV, 
      more than 20 hits in the central jet chamber used to calculate
      the specific energy loss d$E$/d$x$, and the innermost hit of the track 
      must be inside a radius of 60~cm with
      respect to the $z$ axis. The distance of closest approach to 
      the origin must be less than 20~cm in the $z$ direction and less than 1
      cm in the $r\phi$ plane.
\item To reduce background from \epem~annihilation events with $\DST$ mesons
      in the final state,
      the sum of all energy deposits in the ECAL is required to be less 
      than~40 GeV.
      Calorimeter clusters have to pass an energy threshold of 100~MeV for the
      barrel section and 250~MeV for the endcap sections.
\item To reduce the \epem~annihilation background further, the visible 
      invariant mass of the event, $W_{\rm vis}$, 
      should be less
      than $60~{\rm GeV}$. $W_{\rm vis}$ is calculated using the
      energies and positions of clusters measured in the ECAL, the HCAL, 
      the FD and the SW calorimeters and using 
      the momenta of tracks. A matching algorithm~\cite{bib-matching}
      is applied to avoid double-counting of particle momenta in
      the calorimeters and in the tracking chambers.
\item Anti-tagged events are selected by vetoing all events containing
      an energy deposit of more than  50~GeV in the SW or FD in either
      hemisphere of the detector.
      This corresponds to a maximum allowed scattering angle of
      the beam electrons of $\theta'=33~{\rm mrad}$ for electrons
      with $E'>50~{\rm GeV}$.  
\end{itemize}
 
The method of reconstructing $\DST$ mesons is similar
to that used in former OPAL analyses~\cite{bib-opaldst1}. 
It exploits the small mass difference between the $\DST$ and the
$\D0$ mesons which causes the kinetic energy of the slow pion in the decay
$\DSTP \to \D0 \pi^+$ to be only 6~MeV in the $\DST$ rest frame. 
Thus, the combinatorial background is small due to the limited phase space.  
The $\D0$ mesons are identified via their decay $\D0 \to {\rm K}^-
\pi^+$ and $\D0 \to {\rm K}^-\pi^+\pi^-\pi^+$, which form, together
with the slow pion, the ``3-prong'' and ``5-prong'' decay modes
of the $\DST$, respectively.

In the 3-prong decay mode, all combinations of two oppositely
charged tracks in an event are used to form $\D0$ candidates. 
The ${\mathrm d}E/{\mathrm d}x$ probability 
${\cal P}_{\rm K}^{{\rm d}E/{\rm d}x}$ for the kaon hypothesis should
exceed 10$\%$ for at least one of the two tracks. 
The invariant mass $M_{\D0}^{\rm cand}$ of this combination is 
calculated, assigning the kaon mass to the kaon candidate and the pion mass 
to the other track. If for both tracks 
${\cal P}_{\rm K}^{{\rm d}E/{\rm d}x}$ is greater than 10$\%$, 
both possible K$\pi$ combinations are used.   

If $M_{\D0}^{\rm cand}$ lies within a window around
the nominal $\D0$ mass, 
\begin{equation}
1790~{\rm MeV} < M_{\D0}^{\rm cand} < 1940~\rm{MeV},
\end{equation}
the combination is retained as a $\D0$ candidate.
All remaining tracks of opposite charge to the kaon candidate are
then examined
and the invariant mass $M_{\DST}^{\rm cand}$ of the $\DST$ candidates 
is calculated assigning the pion mass to the third track.

Random combinations of low-momentum tracks are
the largest source of background passing the above cuts.
To reduce this background, we exploit the fact that the $\D0$ is
a pseudo-scalar particle which decays isotropically in its rest frame.
This leads to a flat distribution of $\cos\theta^{\ast}$, where
$\theta^{\ast}$ denotes the decay angle between the direction of the
kaon in the $\D0$ rest frame and the direction of the $\D0$ in the
laboratory frame. In contrast, background events exhibit a 
pronounced peak at $\cos\theta^{\ast} = 1$. Therefore we require 
$\cos\theta^{\ast} < 0.9 $.  

In the 5-prong decay mode, the procedure is similar but, due to higher
combinatorial background, some cuts are tightened. To form the $\D0$ candidate,
four tracks are combined
if the charges of the tracks add up to zero. One track
should be identified as a kaon, 
i.e. ${\cal P}_{\rm K}^{{\rm d}E/{\rm d}x}>10\%$, and 
this track's ${\mathrm d}E/{\mathrm d}x$ probability 
${\cal P}_{\pi}^{{\rm d}E/{\rm d}x}$ for the pion hypothesis
should be less than 10$\%$. 
For the other three tracks, ${\cal P}_{\pi}^{{\rm d}E/{\rm d}x}$
is required to be larger than 0.5$\%$. 
If the mass of the $\D0$ candidate lies in the range
\begin{equation}
1830~{\rm MeV}< M_{\D0}^{\rm cand}<1900~{\rm MeV},
\end{equation}
a fifth track is added, with ${\cal P}_{\pi}^{{\rm d}E/{\rm d}x}>0.5\%$
and a charge opposite to the charge of the
kaon candidate, to form the $\DST$ candidate.

To further reduce the combinatorial background in both decay
modes,
a minimum transverse momentum $\ptdst$ of the $\DST$ of 2~GeV is required.
To ensure that the tracks forming the $\DST$ candidates 
are mostly contained in the tracking chambers, 
the pseudorapidity $\etadst$ of the $\DST$ is required 
to be within
$|\etadst| < 1.5$, with $\etadst= -\ln\tan(\theta/2)$. 
The angle $\theta$ is
the polar angle of the $\DST$ candidate.

In about 30$\%$ (8$\%$) of the events with 5-prong (3-prong) candidates,
more than one
$\DST$ candidate passes the above cuts on $\ptdst$ and $\etadst$. 
Since the probability
to correctly reconstruct two different $\DST$ mesons in one event is negligibly
small, only the $\DST$ candidates with $M_{\D0}^{\rm cand}$
closest to the $\D0$ mass of $1864.6~{\rm MeV}$~\cite{bib-PDG}
are retained in events with more than one $\DST$ candidate.
It has been checked that
this method does not produce any biases.
In approximately 6$\%$ of the events,
two or more $\DST$ candidates in an event share the 
same $\D0$ candidate, but different tracks were assigned as slow-pion
candidate. All of these $\DST$ candidates are kept. 

\begin{figure}[htbp]
   \begin{center}
      \mbox{
          \epsfxsize=17.0cm
          \epsffile{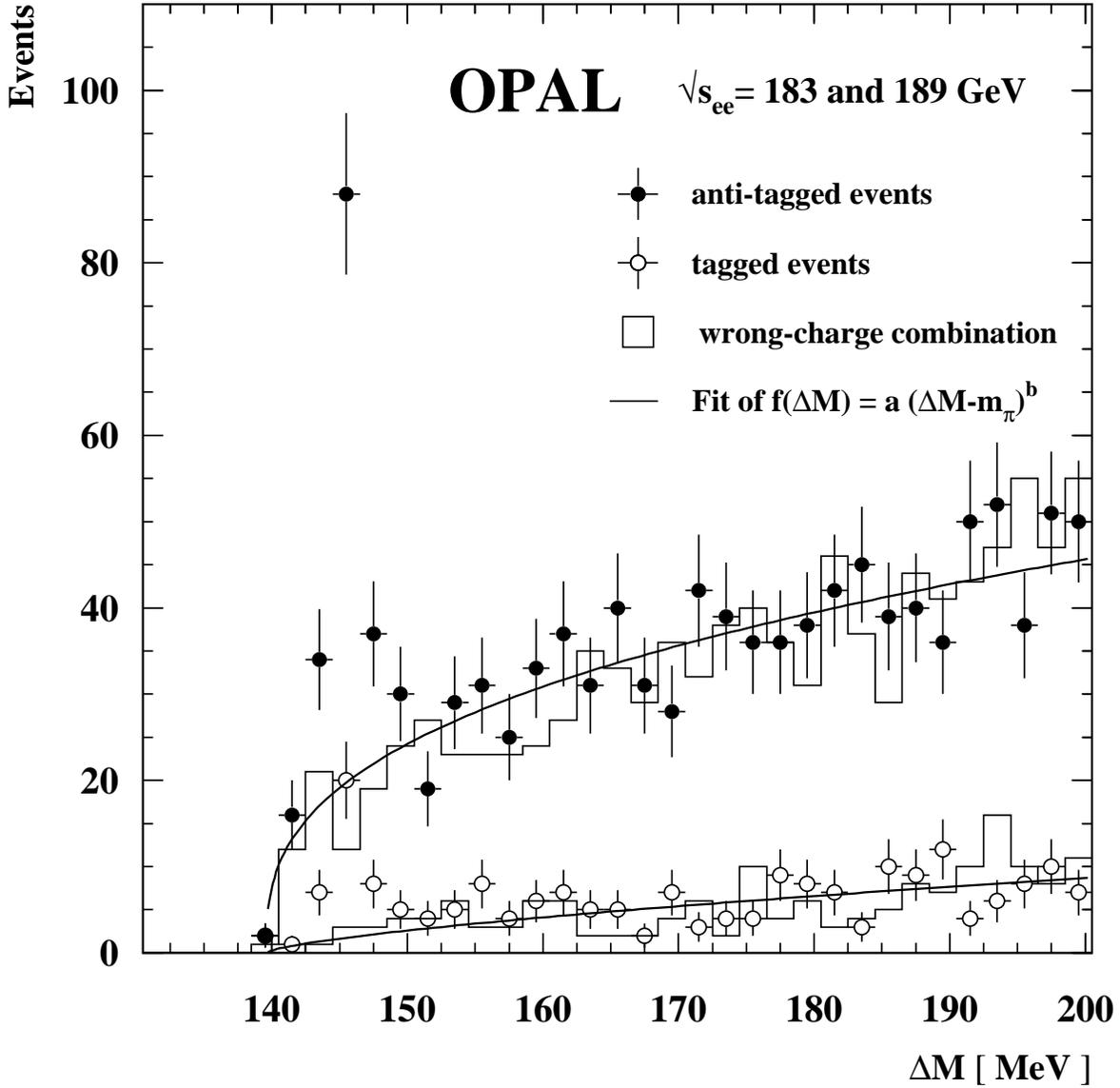}
           }
   \end{center}
\caption{           Mass difference $\DELTACAND$ for both decay modes 
                    for the anti-tagged and tagged sample.
                    In both samples, a clear peak is visible around
                    $\DELTAM = 145.4~{\rm MeV}$. The result of a fit of
                    the background function 
                    $f(\Delta M)=a \cdot (\Delta M-m_{\pi})^b$ to the 
                    upper sidebands is superimposed.
                    The fit regions are $\Delta M>160.5$~MeV for the 
                    anti-tagged
                    events and $\Delta M>154.5$~MeV for the tagged events.
                    The open histograms represent the corresponding 
                    wrong-charge
                    background samples which give a good description 
                    of the combinatorial background.}  
                        
\label{fig-signal}
\end{figure}

Fig.~\ref{fig-signal} shows the difference between the  
$\DST$ and the $\D0$ candidate mass for both decay channels for events with 
$\DELTACAND < 200.5~{\rm MeV}$. 
A clear peak is
observed around $\Delta M = 145.4~{\rm MeV}$ which is the 
mass difference between the $\DST$ and the $\D0$ meson~\cite{bib-PDG}.  
A fit of a background function,
\begin{equation}
f(\Delta M)=a \cdot (\Delta M-m_{\pi})^b,
\label{eq-fit}
\end{equation}
is performed to the upper sideband of the signal, 
defined by $160.5~{\rm MeV}<\Delta M<200.5~{\rm MeV}$, where $m_{\pi}$ is
the pion mass and $a$ and $b$ are free parameters. 
The $\chi^2$
of the fit is 13 for 18 degrees of freedom.
The fit result 
is superimposed for the whole $\Delta M$ range.
In the signal region, defined as $142.5~{\rm MeV}<\Delta M<148.5~{\rm MeV}$,
a number of $100.4\pm12.6~({\rm stat})~\DST$ mesons is obtained 
after subtracting the 
fitted background from the total number
of events in the signal region.
The distribution of 
the wrong-charge background is also shown. 
It is obtained from the data applying identical cuts as for
the signal, but requiring
that the charges of the tracks forming the $\D0$ candidate 
should add up to $-2$ instead of 0. 
In addition, in the 5-prong mode the three pion tracks  
should not have equal charges. 
In the upper sideband, the wrong-charge sample gives a good 
description of the shape and normalisation of the
background in the signal sample.
Hence, no normalisation is applied
to the wrong-charge sample.

Using the Monte Carlo simulations,
$\DST$ mesons produced in \epem~annihilation events 
are found to contribute only around 1$\%$ to the $\DST$ signal. 
It was checked that the \epem~background
is negligible for all values of $\ptdst$. 
Non-photon-photon four-fermion  
background is also found to be negligible.

\section{Separation of direct and single-resolved events}

We use two different methods to study 
the relative contributions of the direct and single-resolved processes
to the data sample. 
First, we study di-jet events using the method described in more detail in
Ref.~\cite{bib-jet}. 
In di-jet events, two experimental variables can be defined, 
$x_{\gamma}^+$ and 
$x_{\gamma}^-$, which are measures of the photon momenta participating
in the hard interaction. They are calculated using the relation 
\begin{equation}
x_{\gamma}^{\pm} 
= \frac{\Sigma_{\rm jets}(E\pm p_z)}{\Sigma_{\rm hadrons}(E\pm 
p_z)},
\end{equation} 
where $p_z$ is the momentum component
along the $z$ axis of the detector and $E$ is the energy of the
jets or hadrons, respectively. Assuming in the LO picture
that the two jets contain all the decay products of the two charm quarks,
we expect for direct events that the whole energy of the event 
is contained in the two jets, i.e.~$x_{\gamma}^+$ and $x_{\gamma}^-$
are close to 1.
In resolved events, there is also energy outside the two jet
cones due to the photon remnant(s). 
Events where either $x_{\gamma}^+$ or $x_{\gamma}^-$ is much smaller than 1
are expected to originate from single-resolved processes. 
Events where both $x_{\gamma}^+$ and $x_{\gamma}^-$ are much smaller than 1
are expected to originate from double-resolved processes. 
The validity of this expectation has been demonstrated in Ref.~\cite{bib-jet}.

The second method can be used for all $\DST$ event, not just
the di-jet sub-sample. We can 
reconstruct the scaled $\DST$ transverse momentum 
$x_{\rm T}^{\DST}$ which is given
by
\begin{equation}
x_{\rm T}^{\DST}=\frac{2\ptdst}{W_{\rm vis}}.
\end{equation}
If $\ptdst$ is a good estimate of the transverse momentum of the charm 
quarks and if the charm quarks are produced centrally ($\eta=0$), 
the variable $x_{\rm T}^{\DST}$ is equal to $x_{\gamma}^{\pm}$.
This variable is therefore sensitive to the ratio of the direct and the 
single-resolved process. As in the case of the $x_{\gamma}^{\pm}$
distribution, the direct contribution dominates at high values of 
$\xt$, whereas
the single-resolved events are concentrated at small $\xt$,
as predicted by the Monte Carlo. 

In order to reconstruct jets,
a cone jet finding algorithm is applied to the signal events.
As in the calculation of $W_{\rm vis}$, the energy and positions of
all clusters in the ECAL, the HCAL, the FD and the SW calorimeters
and the momenta of all tracks are used in the jet finding after
applying the matching algorithm~\cite{bib-matching} to avoid double 
counting of particle momenta. The cone size 
$R=\sqrt{(\Delta\eta)^2+(\Delta\phi)^2}$ 
is set equal to 1, where
$\eta$ and $\phi$ denote the pseudorapidity and the azimuthal angle,
respectively. The minimum transverse
jet energy $E_{\rm T}^{\rm jet}$ is required to be greater than 
$3~{\rm GeV}$. 
The pseudorapidity of the reconstructed jets must be less than 2.

\begin{figure}[htbp]
\vspace*{-2.cm}
   \begin{center}
      \mbox{
          \epsfxsize=15.0cm
          \epsffile{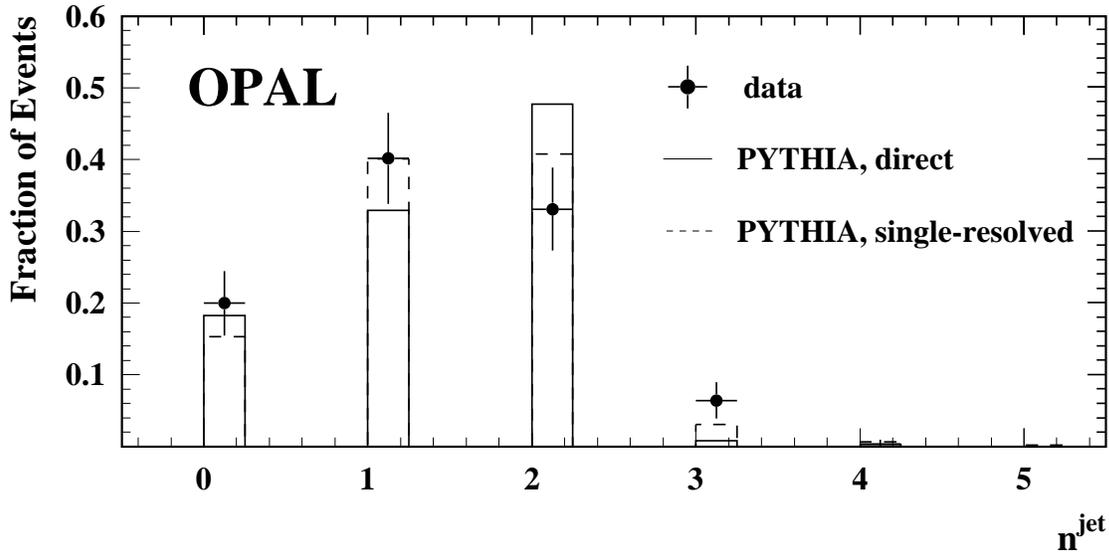}
           }
   \end{center}
\caption{Fraction of signal events with different $n^{\rm jet}$ 
         determined with the cone jet finding algorithm 
         (for anti-tagged events). 
         The dots represent the data after subtraction of
         the combinatorial background. The solid line shows
         the PYTHIA prediction for the direct and the dashed line
         for the single-resolved sample, respectively.} 
\label{fig-njet}
\end{figure}

In Fig.~\ref{fig-njet}, the fraction of events with 
different number of jets, $n^{\rm jet}$, 
is shown for events in the signal region for data and for the direct and the
single-resolved PYTHIA Monte Carlo samples, separately.
About 1/3 of the data events are di-jet events. 
The number of signal events in the direct (single-resolved) Monte Carlo sample is
approximately 6.5 (3.5) times the number of signal events in the data.
In the data, the combinatorial background has
been subtracted using the  $n^{\rm jet}$ distribution from the 
upper $\Delta M$ sideband. The agreement between the $n^{\rm jet}$ 
distributions in the data and the Monte Carlo is satisfactory.
There are slightly more data events in the 3-jet bin and
less data events in the 2-jet bin compared to PYTHIA.
Only a small difference between
the number of jets found in direct and single-resolved events
is expected according to the Monte Carlo.

\begin{figure}[htbp]
   \begin{center}
      \mbox{
          \epsfxsize=15.0cm
          \epsffile{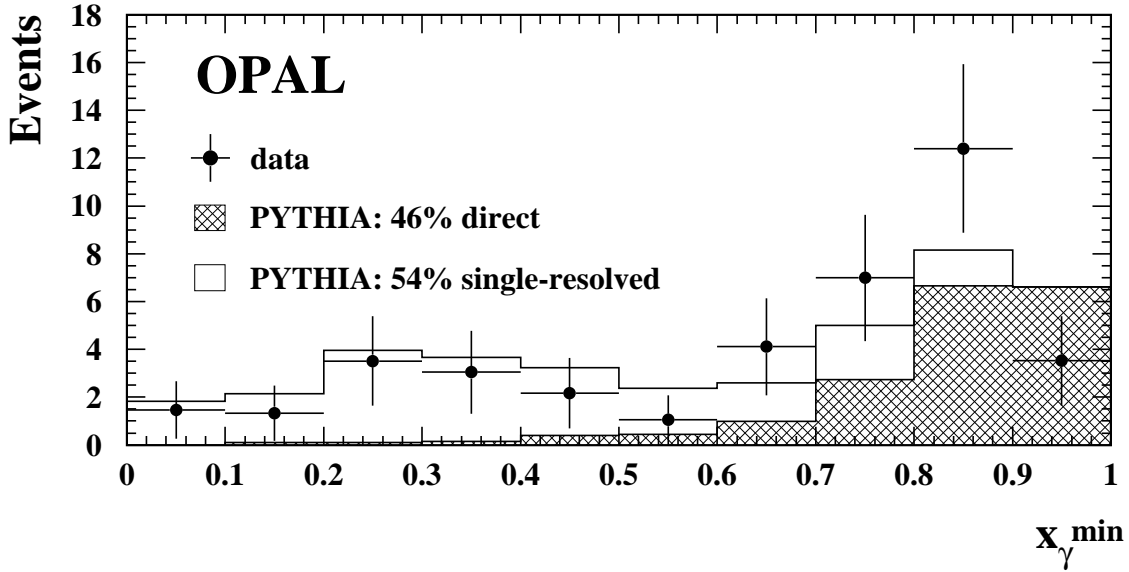}
           }
   \end{center}
\caption{Minimum of $x_{\gamma}^+$ or $x_{\gamma}^-$ for di-jet
         events in the signal region (anti-tagged events only).
         The data, represented by the dots, are background subtracted
         using sideband events. 
         The enhancement at large values of $\xgmin$ is due to the direct
         process. 
         The histograms are the result of a fit of the relative contributions
         of the direct and single-resolved Monte Carlo samples to the data.
         The open histogram shows the single-resolved, the hatched histogram
         the direct contribution to the fit result.}
\label{fig-xgamma}
\end{figure}

Fig.~\ref{fig-xgamma} shows the distribution of   
$\xgmin=\min(x_{\gamma}^+,x_{\gamma}^-)$. 
In the data, the combinatorial background has been subtracted using 
events from the $\Delta M$ sideband. At small values of $\xgmin$,
the distribution is approximately flat, but for large $\xgmin$ values
a clear enhancement is visible. 
This is expected to be due to the direct process. 
The PYTHIA Monte Carlo predicts 87$\%$ of the direct di-jet events to have 
$\xgmin>0.7$ and
82$\%$ of the single-resolved di-jet events to have  $\xgmin<0.7$.

The ratio of direct to single-resolved 
contributions in the di-jet data is determined by a fit 
to the $\xgmin$ distribution using the method of least squares ($\chi^2$ fit).
In the fit, the sum of the direct and
single-resolved Monte Carlo samples is fixed to the number of di-jet events
in the data, but the ratio of direct to single-resolved events is left free.
According to the PYTHIA Monte Carlo, a significant contribution of 
double-resolved events in the data should show up
as a clear enhancement at small $\xgmin$ values. 
As expected from Ref.~\cite{bib-DKZZ}, this is not observed, and
the contribution from double-resolved events is therefore neglected.
The fit yields that $(46\pm11)\%$ of
the di-jet events in the data are due to the direct and $(54\pm11)\%$ are 
due to the single-resolved process. In Fig.~\ref{fig-xgamma}, 
the fitted direct and single-resolved contributions are shown. The $\chi^2$ 
of the fit result is 6.1 for 9 degrees of freedom, and 
the fit result gives a good description of the data. 
At high $\xgmin$ values, where the direct events are concentrated,
the data seem to be slightly shifted
towards smaller $\xgmin$ values compared to the Monte Carlo.

\begin{figure}[htbp]
\vspace*{-3.cm}
   \begin{center}
      \mbox{
          \epsfxsize=15.0cm
          \epsffile{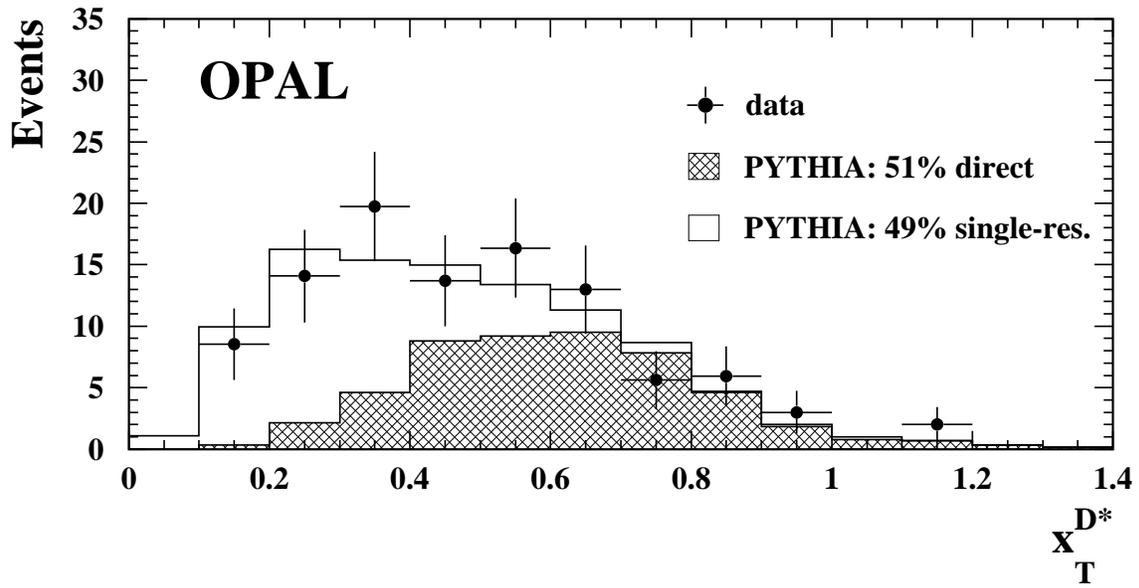}
           }
   \end{center}
\caption{ Scaled transverse momentum of the $\DST$ meson, $\xt$, 
          for all signal events (anti-tagged events only). 
          The dots represent the background subtracted data. 
          The histograms are the result of a fit of the relative 
          contributions of
          the direct and single-resolved Monte Carlo samples to the data. 
          The open histogram shows the single-resolved, 
          the hatched histogram the 
          direct contribution to the fit result.}
\label{fig-ptdst}
\end{figure}

In Fig.~\ref{fig-ptdst}, the scaled $\DST$ transverse momentum $\xt$ 
is plotted for all signal events after subtracting the combinatorial
background. The ratio of direct to 
single-resolved contributions in the data is again determined by a fit 
using the same procedure as applied to the $\xgmin$ distribution. 
Poisson errors are assigned to the bins without data entry.
The fit yields that $(51\pm9)\%$ of
the signal events in the data are due to the direct and $(49\pm9)\%$ are
due to the single-resolved process,  
consistent with the result of the fit to the $\xgmin$ distribution.
In Fig.~\ref{fig-ptdst} the fitted
direct and single-resolved contributions are shown. 
The $\chi^2$ of the fit result is 6.5 for 13 degrees of freedom.
Again, the fit result gives a good description of the data.

Since in the fit to the $\xt$ distribution all signal 
events are used,
whereas in case of the $\xgmin$ distribution the fit is applied only to
a part of the signal events,
we use the result of the fit to the  $\xt$ distribution
in the further analysis.

Table~\ref{tab-eff} summarises the $\DST$ selection efficiencies 
$\epsilon$ for the 3-prong and the 5-prong
decay modes in direct and single-resolved events. 
The efficiencies are calculated using the Monte Carlo by 
dividing the number of reconstructed $\DST$ mesons by
the number of generated $\DST$ mesons with $\ptdst > 2~{\rm GeV}$ 
and $|\etadst| < 1.5$. 
In the 5-prong mode, the selection efficiency for direct 
events is slightly higher than for single-resolved events,
whereas in the 3-prong mode the selection efficiencies are 
about equal for direct and single-resolved event. 
The fractions of direct and single-resolved 
\epem $\to$ \epem $\DST X$ events in the kinematical region 
$\ptdst>2~{\rm GeV}$ and $|\etadst| < 1.5$ are therefore assumed to be
unchanged by the efficiency correction.
To determine the systematic uncertainty, 
half the difference between the direct and single-resolved
efficiencies in the 5-prong mode is used and no systematic uncertainty is used 
for the 3-prong mode.
Together with the relative rate of $\DST$ mesons decaying in the 
5-prong mode and the 3-prong mode, this yields
a 8$\%$ relative uncertainty which 
is added quadratically to the 18$\%$ relative error of the 
direct and single-resolved contributions determined by the fit to the 
$\xt$ distribution.
The direct contribution of the process \epem $\to$ \epem $\DST X$  
in the kinematical region $\ptdst>2$~GeV and $|\etadst|<1.5$ 
is determined to be 
$r_{\rm dir} = (51\pm10)\%$ and the single-resolved process contributes to 
a fraction of $1-r_{\rm dir} = (49\pm10)\%$.
No significant double-resolved contribution is observed.

\section{Differential \boldmath $\DST$ \unboldmath cross-sections}

We determine the differential cross-sections 
${\rm d}\sigma/{\rm d}\ptdst$
and ${\rm d}\sigma/{\rm d}|\etadst|$ for the production
of $\DST$ mesons in anti-tagged \epem$\to$ \epem $\DST X$ events 
as a function of the transverse momentum $\ptdst$ and the 
pseudorapidity $|\etadst|$.
Table~\ref{tab-sigpt} summarises the background-subtracted
number $\ndst$ of $\DST$ mesons and the differential cross-section
${\rm d}\sigma/{\rm d}\ptdst$ for both decay modes.
At large $\ptdst$ the statistical errors are large
for the upper $\Delta M$ 
sidebands of the signal and for the wrong-charge distributions.
These two distributions are therefore combined
to determine the background in each decay mode and $\ptdst$ bin.
The background function $f(\Delta M)=a \cdot (\Delta M-m_{\pi})^b$ 
is fitted to this upper sideband distribution and the number of background
events is calculated from the fit result.

For each decay mode and for each bin in $\ptdst$,  
the differential cross-section $\dspt$ is calculated using the relation 
\begin{equation}
\frac{{\rm d} \sigma}{{\rm d} p_{\rm T}^{\DST}}=
\frac{\ndst}{\epsilon\cdot\BR\cdot{\cal L}\cdot\Delta\ptdst}.
\end{equation}
The efficiency $\epsilon$ is determined using the Monte Carlo by fixing 
the ratio of direct to single-resolved events to the result obtained in the
previous section.
The branching ratios $\BR(\DSTP\to {\rm K}^-\pi^+\pi^+) = 0.02630\pm0.00082$ 
and $\BR(\DSTP\to {\rm K}^-\pi^+\pi^-\pi^+\pi^+)=
0.0519\pm0.0029$ are taken from Ref.~\cite{bib-PDG},
${\cal L}$ is the total integrated luminosity and 
$\Delta\ptdst$ is the width of the $\ptdst$ bin. 
The results of both decay modes agree within the statistical uncertainties. 
The combined differential cross-section $\dspt$ is given in 
Table~\ref{tab-sigpt}. 
The average transverse momentum $\langle\ptdst\rangle$ for each bin
is determined using the method proposed in Ref.~\cite{bib-datapoints}. 

\begin{figure}[htbp]
   \begin{center}
      \mbox{
          \epsfxsize=17.0cm
          \epsffile{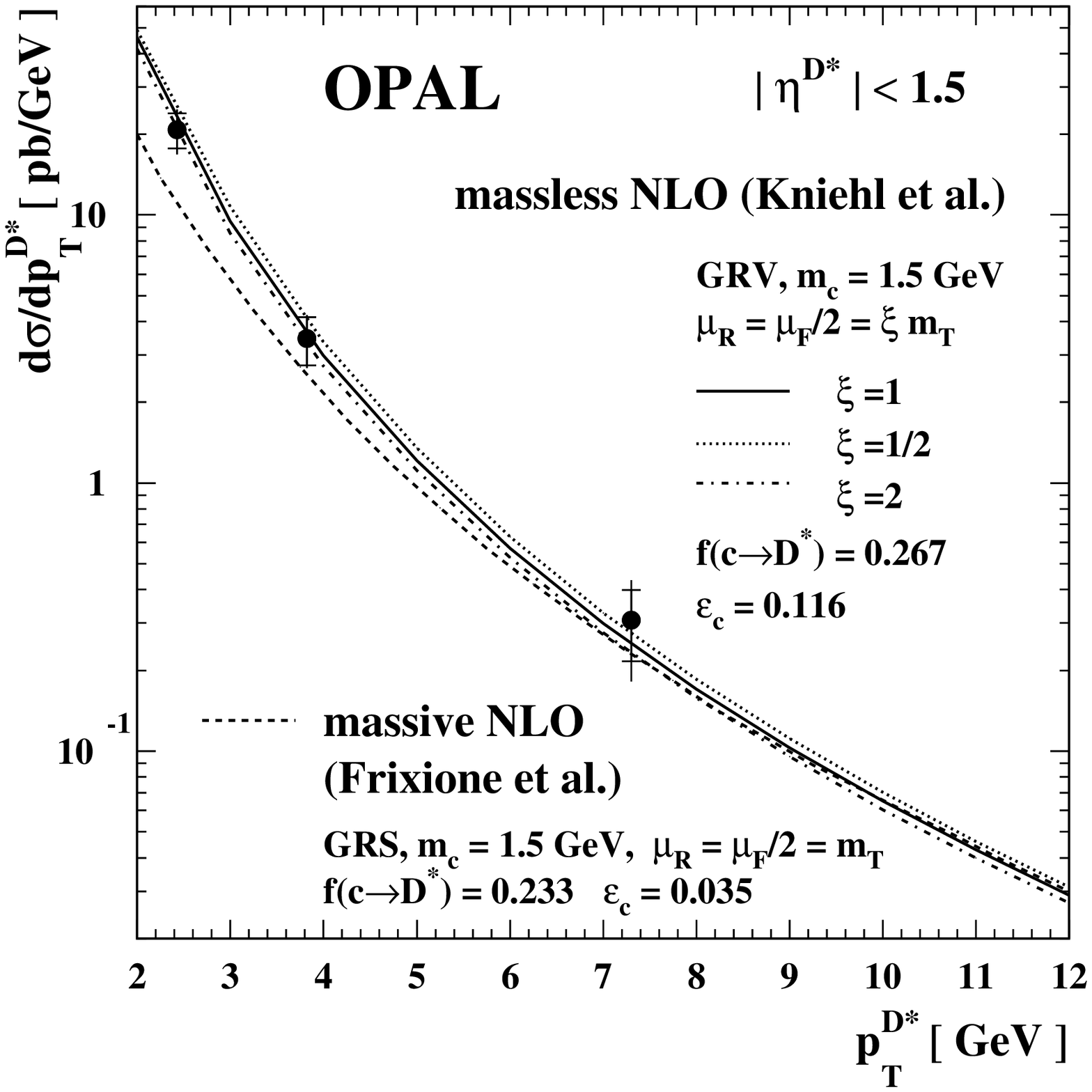}
           }
   \end{center}
\caption{The differential $\DST$ cross-section, $\dspt$, for the process
         \epem $\to$ \epem $\DST X$
         in the range $|\etadst|<1.5$ (for anti-tagged events). 
         The dots represent the combined cross-sections
         from both
         decay modes. The inner error bars give the statistical error and
         the outer error bars the statistical and the systematic error added
         in quadrature.
         The data are compared to a NLO calculation by Kniehl 
         et al.~using the massless approach for three different 
         renormalisation 
         and factorisation scales, $\mu_{\rm R}$ and $\mu_{\rm F}$,
         and to a NLO calculation by Frixione et al. using the
         massive approach. The quantity $m_{\rm T}$ is defined as 
         $m_{\rm T}=\sqrt{p_{\rm T}^2+m_{\rm c}^2}$ where $p_{\rm T}$ is
         the transverse momentum of the charm quark.} 
\label{fig-sigpt}
\end{figure}

 In Fig.~\ref{fig-sigpt}, the combined differential cross-section $\dspt$
 is compared to the NLO calculation by Frixione et 
 al.~\cite{bib-Frixione} using the massive approach and to the NLO calculation
 by Kniehl et~al.~\cite{bib-Kniehl} using the massless approach,
 which was repeated
 by the authors specifically for the kinematical conditions of this analysis.
 In both calculations, the charm quark mass is taken to be 
 $m_{\rm c}=1.5~{\rm GeV}$ and the charm fragmentation is parametrised
 by the Peterson fragmentation function. 
 The Peterson fragmentation parameter $\epsilon_{\rm c}$
 and the fraction $f({\rm c} \to \DSTP)$ of charm
 quarks fragmenting into $\DSTP$ meson are 
 $\epsilon_{\rm c}=0.116$, $f({\rm c} \to \DSTP) = 0.267$ in
 the massless calculation and
 $\epsilon_{\rm c}=0.035$, $f({\rm c} \to \DSTP) = 0.233$
 in the massive calculation.
 For the massless calculation, the parameters were determined via a NLO
 fit~\cite{bib-Kniehl} to LEP1 data on $\DST$ production in \epem~annihilation
 measured by OPAL~\cite{bib-OPALfrag}.
 The renormalisation scale $\mu_{\rm R}$ and the factorisation scale
 $\mu_{\rm F}$ are in both calculations defined as 
 $\mu_{\rm R} = \mu_{\rm F} / 2 = \xi m_{\rm T}$ with  
 $m_{\rm T}=\sqrt{p_{\rm T}^2 + m_{\rm c}^2}$  
 and $\xi =1$, where $p_{\rm T}$ is the transverse momentum of the
 charm quark. 
 The GRV~\cite{bib-grv} parametrisation of the parton distributions of the 
 photon is used in the massless calculation 
 and the GRS~\cite{bib-grs} parametrisation in the massive calculation. 
 Despite the low transverse momenta studied, the agreement between data
 and the massless calculation is good.
 The massive calculation agrees with the data cross-section for
 $\ptdst>3~{\rm GeV}$, but underestimates the data in
 the region of small $\ptdst$. 
 The scale dependence on $\dspt$ determined by using $\xi=1/2$ and $\xi=2$
 is approximately 10$\%$ for both calculations.
 The corresponding curves for the massless case are also shown in 
 Fig.~\ref{fig-sigpt}.
 In addition, the massless calculation was performed using the
 AFG~\cite{bib-afg} and GS~\cite{bib-gs} parametrisations.
 In the massive calculation AFG and GRV were used as
 alternative parametrisations.
 The change of the cross-section is approximately 10$\%$ in both calculations.
 \par
\begin{figure}[htbp]
   \begin{center}
      \mbox{
          \epsfxsize=17.0cm
          \epsffile{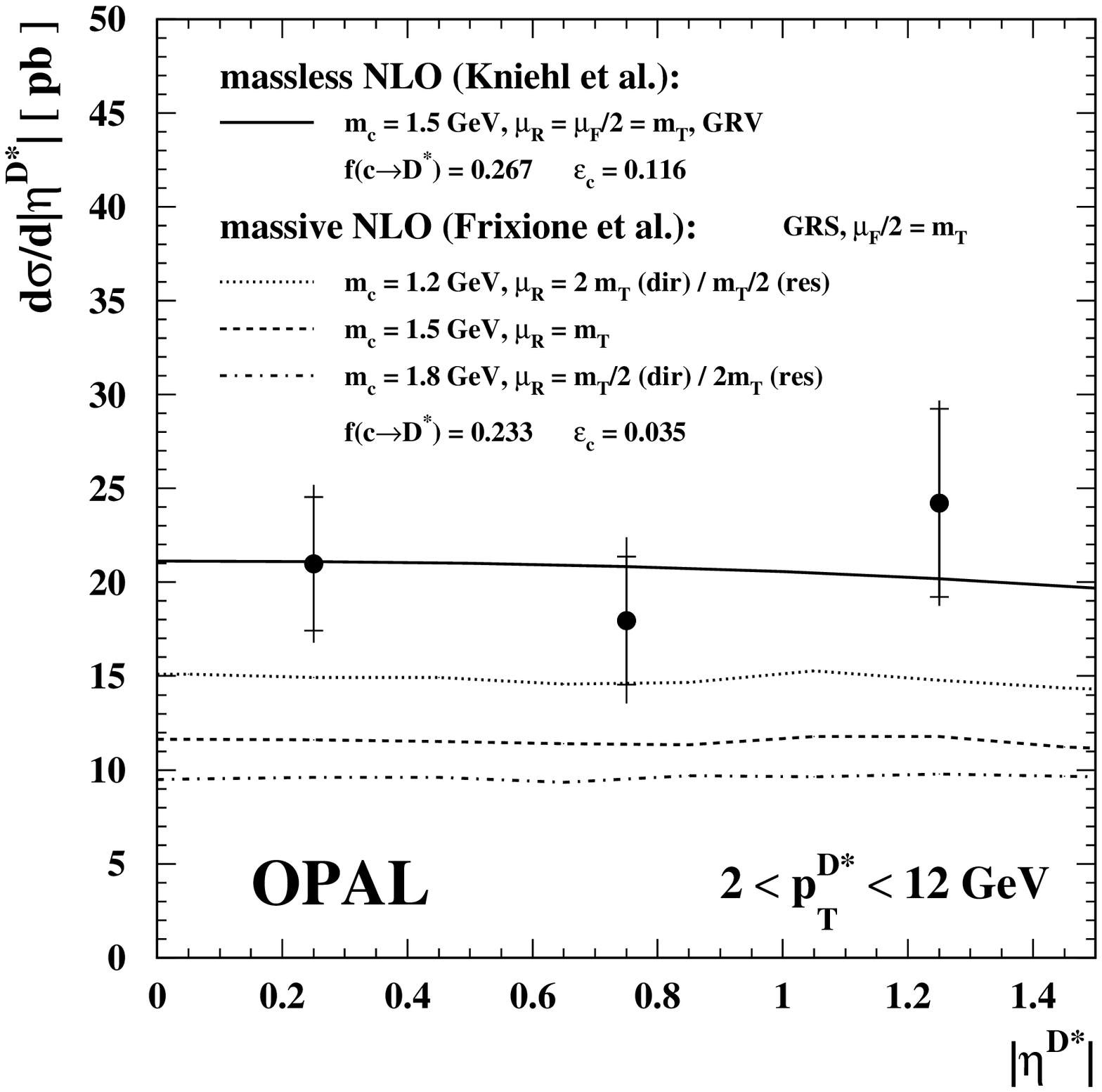}
           }
   \end{center}
\caption{The differential $\DST$ cross-section $\dseta$ for the process 
         \epem $\to$ \epem $\DST X$
         in the range $2~{\rm GeV}<\ptdst<12~{\rm GeV}$ 
         (for anti-tagged events). 
         The dots represent the combined cross-sections
         from both
         investigated decay modes. The inner error bars give 
         the statistical error and
         the outer error bars the statistical and the systematic error
         added in quadrature.
         NLO QCD calculations by Kniehl et~al.~using
         the massless approach are also shown as well as NLO QCD 
         calculations by 
         Frixione et al.~using 
         the massive approach using different renormalisation scales
         separately for the direct (dir) and the single-resolved (res) 
         contributions
         and different charm quark masses. The quantity $m_{\rm T}$ is 
         defined as $m_{\rm T}=\sqrt{p_{\rm T}^2+m_{\rm c}^2}$, 
         where $p_{\rm T}$ is
         the transverse momentum of the charm quark.}
\label{fig-sigeta}
\end{figure}

 In Table~\ref{tab-sigeta}, the number of reconstructed $\DST$ mesons
 with $2~{\rm GeV}<\ptdst<12~{\rm GeV}$ and $|\etadst|<1.5$ and the
 corresponding differential cross-sections $\dseta$ are given as a 
 function of $|\etadst|$ for both decay modes.
 The numbers are determined in the same way as described above in the case
 of $\ptdst$. Within the statistical uncertainties, both decay modes yield
 comparable results.
 The combined differential cross-section is also given in
 Table~\ref{tab-sigeta} and plotted in Fig.~\ref{fig-sigeta}.
 The distribution is dominated by the events at low $\ptdst$, and
 within the error, it is independent of $|\etadst|$.
 The centres of the bins are taken as the average $\langle|\etadst|\rangle$ 
 values. 
 The massless calculation by Kniehl et al.~is in good agreement with the
 measured differential cross-section, whereas the massive calculation
 of Frixione et al.~underestimates the data, as seen already in 
 Fig.~\ref{fig-sigpt}. 
 For the massive calculation, two additional curves are shown in
 Fig.~\ref{fig-sigeta} representing different charm quark masses $m_{\rm c}$ 
 with 
 renormalisation scales $\mu_{\rm R}$ and factorisation scales $\mu_{\rm F}$
 as indicated in the figure.
 The combination of a small charm quark mass ($m_{\rm c}=1.2$~GeV) 
 with a special choice of the renormalisation scale 
 ($\mu_{\rm R}=2m_{\rm T}$ for the direct and 
 $\mu_{\rm R}=m_{\rm T}/2$ for the single-resolved process) yields
 a cross-section which is closer to the data, but still slightly 
 low.

For the determination
of the systematic uncertainties,
each decay mode and each bin in $\ptdst$ or $|\etadst|$
is treated individually. The following errors are taken into account:
\begin{itemize}
\item The relative uncertainties on 
      $\BR(\DSTP\to {\rm K}^-\pi^+\pi^+)$ and 
      $\BR(\DSTP\to {\rm K}^-\pi^+\pi^-\pi^+\pi^+)$ of $3.1\%$ 
      and $5.6\%$, respectively~\cite{bib-PDG}.   
\item The relative uncertainty on the selection efficiencies due to the 
      limited number of Monte Carlo events
      varies between 5$\%$ in the lowest
      $\ptdst$ bin for the 3-prong decay mode and 
      13$\%$ in the highest $\ptdst$ bin 
      for the 5-prong decay mode. The corresponding errors 
      for the $|\etadst|$ distribution are 5$\%$ to 7$\%$ in all 
      $|\etadst|$ bins except for the bin $1<|\etadst|<1.5$
      in the 5-prong mode where the error is 12$\%$.
\item The uncertainty on the number of background events determined 
      from the fit
      of the background function to the sum of the sidebands of the 
      signal data 
      and of the wrong-charge 
      distribution.
      A modified background function is constructed, defined by
      the requirement $\chi^2=\chi^2_{\rm min}+1$,  where $\chi^2_{\rm min}$
      is the minimum $\chi^2$ of the fit.
      The relative difference between the number of background events
      determined with the modified background function and the number
      of background events determined with the original 
      background function is taken as the error. 
      Depending on the decay mode and on the $\ptdst$ or
      $|\etadst|$ bins, the relative uncertainties vary between 
      5$\%$ and 25$\%$. 
\item The contributions of the direct and single-resolved  
      Monte Carlo samples 
      have been varied between $40\%$ and $60\%$. 
      In the 3-prong and in the 5-prong mode, the corresponding errors
      on the cross-sections are smaller than 7$\%$ for all bins in $\ptdst$
      and in $|\etadst|$. 
\item Uncertainties in the modelling of the tracking in the central 
      detector are assessed by repeating the analysis with the tracking
      resolutions varied in the Monte Carlo 
      by $\pm 10\%$ around the values that describe
      the data best. The efficiencies obtained are compared with
      the original values, and the relative difference is quoted as
      the systematic error. Depending on decay mode and bin, 
      this error lies between 5$\%$ and 15$\%$.
\item Uncertainties in the d$E/$d$x$ probabilities for identifying kaons.
      In a former OPAL analysis \cite{bib-dedx},  
      $\DST$ mesons are reconstructed in the 3-prong mode using a similar 
      set of cuts to this analysis.
      The relative error on the d$E/$d$x$ probability for identifying 
      kaons is determined to be around 3$\%$. 
      In the 5-prong mode, not studied in Ref.~\cite{bib-dedx}, 
      the pion tracks are also identified
      using the d$E/$d$x$ probabilities, so the corresponding uncertainty  
      is assumed to be 5$\%$.
\end{itemize}  

The uncertainty on the integrated luminosity ${\cal L}$ is smaller than 
1$\%$ and is therefore not taken into account. 
The individual systematic uncertainties are added quadratically, 
separately for each
decay mode as well as for each bin in $\ptdst$ or $|\etadst|$. 
The integrated cross-section $\sigma_{\rm meas}$ of the
process \epem $\to$ \epem $\DST X$ in the kinematical region
$2~{\rm GeV}<\ptdst<12~{\rm GeV}$ and $|\etadst|<1.5$ is determined to be
$\sigma_{\rm meas}^{\DST} = 29.4\pm3.4({\rm stat})\pm2.4({\rm sys})$~pb.

The LO cross-sections for the direct process, $\sigma_{\rm dir}^{\DST}$,
and for the single-resolved process, $\sigma_{\rm res}^{\DST}$,
calculated with the PYTHIA Monte Carlo for different 
LO parametrisations of the parton densities (SaS-1D~\cite{bib-sas}, 
GRV~\cite{bib-grv} and
LAC1~\cite{bib-lac}) are given in Table~\ref{tab-sigma}.
In the PYTHIA Monte Carlo, the charm quark mass $m_{\rm c}$ was
varied between 1.3 and 1.7 GeV. Since the ratio of direct to single-resolved
cross-sections is about 1:1 in the data, the direct cross-section is
well described by PYTHIA and the single-resolved cross-section
is best described using GRV. The single-resolved cross-section is
underestimated using SaS-1D, and the LAC1 parametrisation
overestimates the single-resolved cross-section.

\section{\boldmath Total cross-section $\sigcc$ \unboldmath}   

For the determination of the total cross-section $\sigdst$, the Monte Carlo
is used to extrapolate to the full kinematical region using the relation
\begin{eqnarray}
\sigdst & = & \sigdst_{\rm dir}~+~\sigdst_{\rm res}\nonumber\\
        & = & \rule[0.cm]{0.cm}{.7cm}\sigma_{\rm meas} \left( r_{\rm dir} \cdot {\cal R_{\rm dir}^{\rm MC}}
          + (1-r_{\rm dir}) \cdot {\cal R_{\rm res}^{\rm MC}} \right).
\label{eq-sigdstar}
\end{eqnarray}
where ${\cal R_{\rm dir}^{\rm MC}}$ and ${\cal R_{\rm res}^{\rm MC}}$ are
the extrapolation factors. This allows
the total cross-section of the process \epem $\to$ \epem
$\rm{c\bar{c}}$ for $Q_i^2<4.5~{\rm GeV}^2$ to be calculated using the equation
\begin{equation}
\sigcc = \frac{1}{2\cdot f({\rm c} \to \DSTP)} \cdot \sigdst.
\label{eq-sigcc}
\end{equation}
In a previous publication~\cite{bib-opaldst3}, the product $P_{\rm c}=
f({\rm c}\to\DSTP)\times\BR(\DSTP\to \D0 \pi^+)\times \BR(\D0 \to 
{\rm K}^- \pi^+)$ 
was derived from measurements of $\DST$ production in \epem collisions at 
$\sqee = 10.5$ and $30$~GeV to be $P_{\rm c}=(7.1\pm0.5)\cdot 10^{-3}$. 
With the branching ratios taken
from Ref.~\cite{bib-PDG}, a hadronisation fraction 
$f({\rm c} \to \DSTP) = 0.270\pm0.019\pm0.010$ is derived, where the last error
is due to the branching ratio uncertainties.
Since the invariant mass range
of the photon-photon system studied in this analysis is of the same order
of magnitude as
the \epem energies mentioned above, this value of
$f({\rm c}\to \DST)$ is used in the analysis.

The extrapolation factors ${\cal R_{\rm dir}^{\rm MC}}$ 
for the direct events and ${\cal R_{\rm res}^{\rm MC}}$ for
the single-resolved events are
defined as the ratio of the number of all generated $\DST$ mesons
in the full kinematic range of $\ptdst$ and $|\etadst|$
divided by the number of generated $\DST$ mesons with
$2~{\rm GeV}<\ptdst<12~{\rm GeV}$ and $|\etadst|<1.5$.  
The extrapolation factors are 
${\cal R_{\rm dir}^{\rm MC}}=12.6$ and 
${\cal R_{\rm res}^{\rm MC}}=18.4$ 
obtained using the combination $m_{\rm c}=1.5~{\rm GeV}$ and the 
Peterson fragmentation with $\epsilon_{\rm c}=0.031$. 

The extrapolation introduces systematic uncertainties 
due to the modelling of the fragmentation of the 
charm quarks into $\DST$ mesons which influence mainly the $\ptdst$
distributions. 
To determine the systematic errors on the 
extrapolation factors,
different $m_{\rm c}$ values and different fragmentation functions were used
for the event generation in the Monte Carlo: 
\begin{itemize}
\item The charm quark mass $m_{\rm c}$ was
varied between 1.3 and 1.7~GeV. 
\item $\epsilon_{\rm c}=0.0851$~\cite{bib-Kniehl}
was used in the Peterson fragmentation function.
\item The Lund symmetric
fragmentation function~\cite{bib-lundff} was used with the parameters
$a=1.95$ and $b=1.58$ determined in Ref.~\cite{bib-opaldst3}.
\item In case of the single-resolved process,
the GRV parametrisation was used as an alternative parametrisation
in combination with different charm quark masses and fragmentation functions. 
\end{itemize}

For all studied combinations of $m_{\rm c}$, fragmentation functions
and parametrisations of the parton densities,
the direct and the single-resolved Monte Carlo samples were added
in such a way that 
in the kinematical range $2~{\rm GeV}<\ptdst<12~{\rm GeV}$ and $|\etadst|<1.5$ 
the cross-section is equal
to $\sigma_{\rm meas}$, the direct contribution is $r_{\rm dir}=51\%$
and the single-resolved contribution is $1-r_{\rm dir}=49\%$. 
The resulting differential cross-sections 
${\rm d}\sigma/{\rm d}p_{\rm T}^{\DST}$ and
${\rm d}\sigma/{\rm d}|\eta^{\DST}|$ are in agreement  
with the measured differential cross-sections 
(Figs.~\ref{fig-sigpt}-\ref{fig-sigeta}). 
Therefore the mean quadratic deviation of all 
calculated extrapolation factors from the central values 
${\cal R_{\rm dir}^{\rm MC}}=12.6$ and 
${\cal R_{\rm res}^{\rm MC}}=18.4$ 
is used to determine the relative systematic uncertainties of $19\%$
for ${\cal R_{\rm dir}^{\rm MC}}$ and $27\%$ for 
${\cal R_{\rm res}^{\rm MC}}$. 

Combining Eqs.~(\ref{eq-sigdstar}) and (\ref{eq-sigcc}), we determine
the total cross-section of the process \epem $\to$ \epem $\rm{c\bar{c}}$ to
be $\sigcc = \sigtot$ at $\sqee=183-189$~GeV.
The first error is the statistical, the second error is
the systematic error
and the third error is the extrapolation uncertainty. 
The direct contribution is determined to be 
$\sigcc_{\rm dir} = \sigdir$ and the single-resolved contribution
to be $\sigcc_{\rm res} = \sigres$.

The separation of direct and resolved events in heavy
quark production is scheme-dependent in the NLO massless
calculation, but it is unambiguous in LO and in the NLO massive calculation.
Using $m_{\rm c}=1.5$~GeV, 
the LO direct cross-section in PYTHIA lies in the range $300^{+41}_{-44}$~pb.
The LO calculation of Ref.~\cite{bib-DKZZ} gives $382^{+186}_{-94}$~pb,
and the NLO calculation $593^{+319}_{-198}$~pb. The upper and lower
error correspond to $m_{\rm c}=1.3$ and $1.7$~GeV, respectively.
The OPAL anti-tagging condition was applied to these calculations.
The measured direct cross-section agrees well with the LO
calculations, whereas it lies at the lower end of the NLO calculation. 
This could be due to the
separation procedure for direct and single-resolved events 
which uses distributions from a LO Monte Carlo.

\begin{figure}[htbp]
   \begin{center}
      \mbox{
          \epsfxsize=17.0cm
          \epsffile{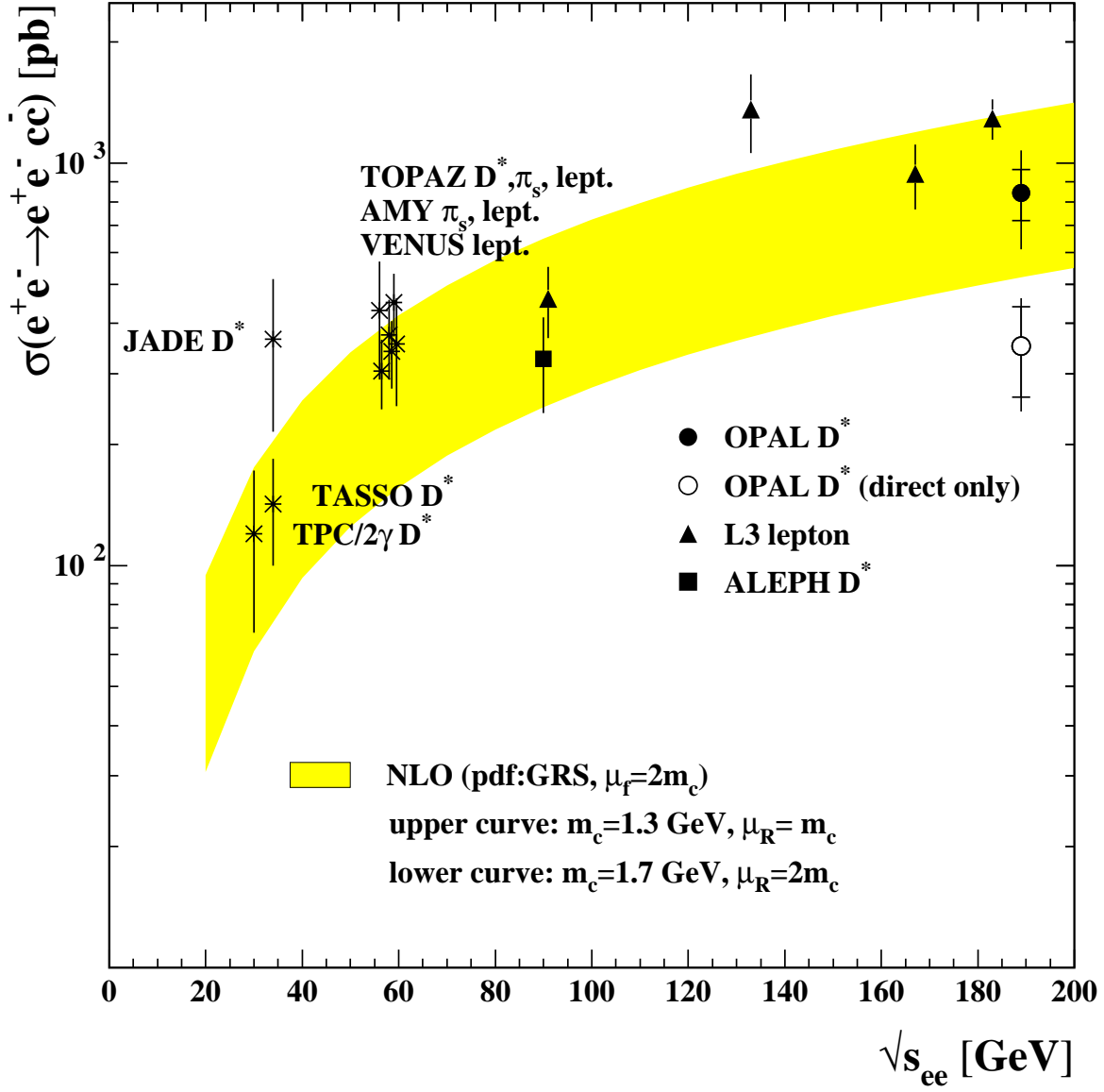}
           }
   \end{center}
\caption{Comparison of measured cross-sections for the process
         \epem $\to$ \epem $\ccbar$ where the charm quarks are
         produced in the collision of two quasi-real photons.
         The outer error bars on the OPAL points represent the
         total errors, including the extrapolation uncertainty, and
         the inner bars are the statistical errors.
         The values for TASSO, TPC/2$\gamma$,
         JADE, TOPAZ, AMY and VENUS are taken from 
         Ref.~\protect\cite{bib-LEP2},
         for ALEPH from Ref.~\protect\cite{bib-ALEPH} and for 
         L3 from Ref.~\protect\cite{bib-L3}. 
         The band shows a NLO calculation 
         of the process \epem $\to$ \epem $\ccbar$~\protect\cite{bib-DKZZ} 
         for a charm quark mass between 1.3 and 1.7~GeV using 
         the GRS parametrisation for 
         the parton distributions of the photon.} 
\label{fig-sigtot}
\end{figure}

Fig.~\ref{fig-sigtot} shows the total cross-section $\sigcc$ 
compared to other measurements and
to the NLO calculation of Ref.~\cite{bib-DKZZ} using
the GRS parton distributions.
The calculation is in
good agreement with the OPAL result within the large
band of uncertainties due to variations of $m_{\rm c}$,
$\mu_{\rm R}$ and $\mu_{\rm F}$. 
The anti-tagging condition used in this paper has been
applied to the NLO calculation. It should be noted that
the anti-tagging conditions of the different experiments
are not identical.
The OPAL result is consistent with the L3 result
at $\sqee=167$~GeV and about 1.5 standard deviations below
the L3 result at $\sqee=183$~GeV~\cite{bib-L3}.

\section{Determination of \boldmath $\CF2$ \unboldmath}
%
 In this section, deep inelastic electron-photon scattering is studied
 using single-tagged events.
 The D* production cross-section, as well as the charm 
 production cross-section and the charm structure function
 $\CF2$ of the photon are determined from events with a beam
 electron scattered into the forward detectors (tagged events).
 An event is tagged (SW-tagged or FD-tagged) if the energy of 
 the scattered electron $E'$, measured in the angular range
 $33~{\rm mrad} < \theta' < 55~{\rm mrad}$ for the SW or 
 $60~{\rm mrad} < \theta' < 120~{\rm mrad}$
 for the FD, exceeds 50~GeV in one hemisphere of the detector. 
 The corresponding approximate ranges in $Q^2$ are 
 $ 5~{\rm GeV}^2<Q^2< 30~{\rm GeV}^2$ and
 $30~{\rm GeV}^2<Q^2<100~{\rm GeV}^2$ 
 for SW-tagged and FD-tagged events, respectively.
 The selection of $\DST$ candidates is identical to the selection  
 in the anti-tagged case, with three exceptions: 
\begin{itemize}
\item For the calculation of the visible invariant mass $W_{\rm vis}$, 
      clusters in the SW or FD in the hemisphere of the tag are excluded. 
\item 
      The combinatorial background in tagged events is smaller due to
      the slightly smaller mean number of tracks per event.
      This makes it possible to include $\DST$ mesons with
      $\ptdst>1$~GeV for SW-tagged events.
\item In FD-tagged events, the $\DST$ mesons have higher
      transverse momenta $\ptdst$ due to the transverse momentum balance 
      between
      the tagged electron and the hadronic system. To improve the signal
      to background ratio,
      a cut $\ptdst>3~{\rm GeV}$ is applied for FD-tagged events.
\end{itemize}
 Fig.~\ref{fig-signal} shows the distribution of the difference between
 the $\DST$ and the $\D0$ candidate mass found in the tagged sample.
 The fit of the
 background function Eq.~\ref{eq-fit} to the upper sideband of the signal
 was performed in the range
 $154.5~{\rm MeV}<\Delta M<200.5~{\rm MeV}$.
 The $\chi^2$ of the fit result is 25 for 21 degrees of freedom.
 Subtracting the background predicted by the fit, $\ntag$ 
 $\DST$ mesons are found in the signal region of the tagged events.
 The combinatorial background in 
 the upper sideband is also well described by the tagged wrong-charge sample.
 Background subtraction with the wrong-charge sample gives a consistent result
 for the number of $\DST$ events.
 Due to the small number of $\DST$ mesons both $\DST$ decay
 modes are combined for the further analysis. 
 No double-tagged $\DST$ event has been found, i.e. 
 an event with energy deposits of more than
 50~GeV in the forward calorimeters in both hemispheres.

\begin{figure}[htbp]
   \begin{center}
      \mbox{
          \epsfxsize=16.0cm
          \epsffile{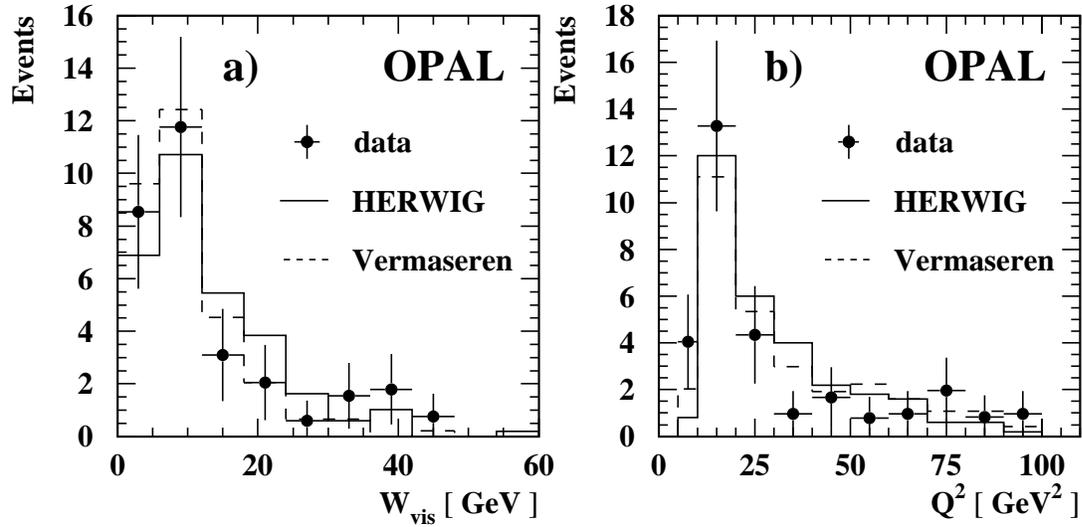}
           }
   \end{center}
\caption{The distributions of a) the visible invariant mass, $W_{\rm vis}$,
         and b) the negative four-momentum squared, $Q^2$, for the tagged 
         signal events.
         The data are compared to the predictions of the HERWIG and Vermaseren
         generators. The Monte Carlo distributions are normalised
         to the number of data events.}
\label{fig-wvis}
\end{figure}

\begin{figure}[htbp]
   \begin{center}
      \mbox{
          \epsfxsize=15.0cm
          \epsffile{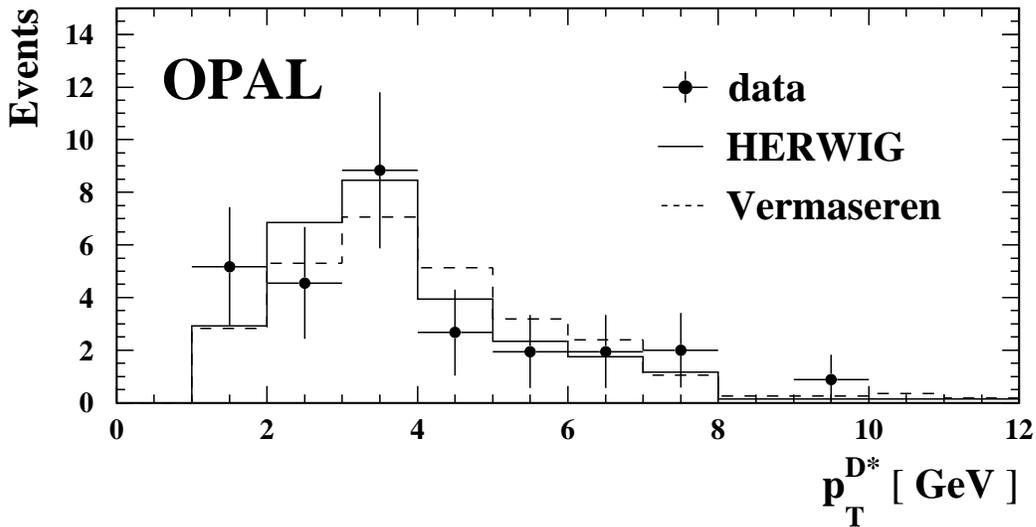}
           }
   \end{center}
\caption{The distribution of transverse momentum $\ptdst$ for the
         $\DST$ mesons in the tagged signal events.
         The data are compared to the predictions of the HERWIG and Vermaseren
         generators. The Monte Carlo distributions are normalised
         to the number of data events.}
\label{fig-pttag}
\end{figure}

 Fig.~\ref{fig-wvis} shows the distributions of $W_{\rm vis}$ and
 of the measured $Q^2$ of the tagged signal events, and 
 the $\ptdst$ distribution is presented in Fig.~\ref{fig-pttag}.  
 The data are compared to the predictions of the HERWIG and Vermaseren
 Monte Carlo generators, normalised to the number of data events. 
 Both Monte Carlo generators give a good description of the shape of the
 data distributions.

 We determine the cross-section for $\DST$ production in deep inelastic
 electron-photon scattering in the well-measured kinematic range:
 $\ptdst>1~{\rm GeV}$ for an electron scattering angle 
 $33~{\rm mrad}<\theta'<55~{\rm mrad~(SW)}$ 
 or $\ptdst>3~{\rm GeV}$ for $60~{\rm mrad}<\theta'<120~{\rm mrad~(FD)}$,
 $|\etadst|<1.5$ and $E'>50~{\rm GeV}$,
 $5~{\rm GeV}^2<Q^2<100~{\rm GeV}^2$; using almost the whole accessible
 $Q^2$ range defined by $\theta'$ and $E'$.

 The analysis is performed in two bins of $x$ with $0.0014<x<0.1$
 and $0.1<x<0.87$.
 The $x$ range is limited by the $Q^2$ range, by the minimum kinematically
 allowed invariant mass  $W>3.88~{\rm GeV}$ needed to produce a $\DST$ meson,
 and by the event selection cut $W_{\rm vis}<60$~GeV.
 To take into account the detector acceptance and resolution in $x$
 the data are corrected using a $2\times2$~matrix.
 The measured $x_{\rm vis}$ is calculated from Eq.~\ref{eq-x}
 using $W_{\rm vis}$ and the measured value of $Q^2$. The resolution
 effects in the $Q^2$ reconstruction are small compared to the
 resolution effects in measuring $x$. They can therefore be neglected.

 The $\DST$ selection efficiency for $x>0.1$ is given by the ratio
 of the number of reconstructed $\DST$ mesons originating 
 from events with $x>0.1$ to all generated $\DST$ mesons 
 in events with $x>0.1$, in the restricted kinematic range defined above. 
 The selection efficiency for $x>0.1$ is about $(21\pm2)\%$ (not 
 including the branching ratios).
 For $x>0.1$ the selection efficiencies obtained from
 both Monte Carlo generators are consistent, but for $x<0.1$
 the selection efficiencies are around $(30\pm 3)\%$ according to 
 HERWIG and around $(18\pm 2)\%$ according to Vermaseren. 
 Both programs predict that about one third of the selected $\DST$ events  
 generated with $x<0.1$ are reconstructed with $x_{\rm vis}>0.1$, whereas
 migration from $x>0.1$ to $x_{\rm vis}<0.1$ is very small. 

 Table~\ref{tab-unfold} summarises the number of reconstructed $\DST$
 mesons and gives the measured values of the cross-section
 \begin{equation} 
 \sigma_{\rm tag}^{\DST}=
 \frac{N_{\DST}^{\rm cor}}{{\rm BR} \cdot {\cal L}},
 \end{equation}
 which is the deep inelastic
 electron-photon scattering cross-section for $\DST$ production in 
 the restricted kinematic range as defined above.
 It is calculated from the number of $\DST$ events,
 $N_{\DST}^{\rm cor}$, obtained from the $2\times 2$ matrix correction 
 using both HERWIG and Vermaseren.  
 For the combined $\DST$ branching ratios into the 3-prong and 
 5-prong mode, we use $\BR=0.0782\pm0.0030$~\cite{bib-PDG}.
 The total integrated luminosity ${\cal L}$ is 220~pb$^{-1}$.
 The average of the cross-sections corrected with HERWIG and Vermaseren
 is also given in Table~\ref{tab-unfold}.

 For $x>0.1$, both Monte Carlo models yield consistent results.
 For $x<0.1$, the difference between the cross-sections
 $\sigma_{\rm tag}^{\DST}$ obtained using HERWIG and 
 Vermaseren is due to the different $\DST$ selection efficiencies. 
 The following systematic errors on $\sigma_{\rm tag}^{\DST}$ are taken into
 account:
\begin{itemize}
\item The limited number of Monte Carlo events leads to
      an uncertainty of approximately 15$\%$ on 
      $\sigma_{\rm tag}^{\DST}$ 
      for each of the Monte Carlo generators. 
\item
      Within the statistical uncertainties, the HERWIG and Vermaseren models 
      yield consistent corrected numbers of $\DST$ mesons for $x>0.1$, whereas
      for $x<0.1$, the corrected numbers of $\DST$ mesons obtained with
      the HERWIG and Vermaseren Monte Carlo models differ by more than
      2 standard deviations. Therefore only for $x<0.1$, half the difference
      between $\sigma_{\rm tag}^{\DST}$ using HERWIG
      and Vermaseren is taken as error on the averaged value 
      of $\sigma_{\rm tag}^{\DST}$.
\item The combined relative uncertainty on the branching ratios
      $\BR(\DSTP\to {\rm K}^-\pi^+\pi^+)$ and
      $\BR(\DSTP\to {\rm K}^-\pi^+\pi^-\pi^+\pi^+)$ is $3.8\%$~\cite{bib-PDG}.

\item The uncertainty in the number of background events in the
      signal region estimated in the same way as for anti-tagged events
      gives relative errors of 12$\%$ for $x<0.1$ and 6$\%$ for $x>0.1$. 
\item The relative uncertainty due to the modelling of the tracking in 
      the central detector is estimated to be 8$\%$ using the corresponding
      errors determined for the differential cross-sections
      $\dspt$ and $\dseta$ for the anti-tagged events.
\item The relative uncertainty due to the use of the 
      d$E/$d$x$ probabilities for identifying kaons and
      pions is estimated to be 4$\%$, also using the corresponding
      errors determined for the differential cross-sections
      $\dspt$ and $\dseta$ for the anti-tagged events. 
\item The uncertainty due to the measurement of the energy $E'$  
      of the tagged electron is assessed by 
      shifting the reconstructed quantity in the Monte Carlo
      according to its resolution and by repeating the  
      analysis. The change on 
      $\sigma_{\rm tag}^{\DST}$  
      is around 3$\%$ and is taken into account as relative error. 
      The uncertainty due to the measurement of the scattering angle $\theta'$ 
      is determined in the same way as for $E'$.
      The relative change on $\sigma_{\rm tag}^{\DST}$ 
      is found to be only around 1$\%$. This error is therefore neglected.
      The uncertainty due to the measurement of the visible invariant
      mass $\Wvis$ of the event is estimated to be only around 1$\%$ and
      is therefore also neglected. 
\end{itemize}
 All systematic errors are added in quadrature.
 \par
 For the determination of the total cross-section of $\DST$ production in
 deep inelastic electron-photon scattering, $\sigdst$,  
 the Monte Carlo models are used to extrapolate to the whole kinematic region.
 This allows the total charm cross-section in deep
 inelastic electron-photon scattering to be calculated via the relation 
\begin{eqnarray}
\sigcc  & = & \frac{1}{2\cdot f({\rm c} \to \DSTP)} \cdot \sigdst \nonumber\\
        & = & \rule[0.cm]{0.cm}{.7cm} \frac{1}{2\cdot f({\rm c} \to \DSTP)} 
             \cdot{\cal R_{\rm tag}^{\rm MC}\cdot\sigma_{\rm tag}^{\DST}}.
\label{eq-sigtag}
\end{eqnarray}
 The extrapolation factor ${\cal R}_{\rm tag}^{\rm MC}$ is defined in the
 same way as in the anti-tagged case.
 Table~\ref{tab-tagda} gives the values of the 
 total charm cross-section, $\sigcc$,
 extrapolated using HERWIG and Vermaseren as well as
 the averaged cross-section. The extrapolation error has been determined
 in the following way: 

 For $x>0.1$, both Monte Carlo generators predict very similar values for 
 ${\cal R}_{\rm tag}^{\rm MC}$ (4.8/4.6 for HERWIG/Vermaseren)
 and thus for $\sigcc$.
 The uncertainty on ${\cal R}_{\rm tag}^{\rm MC}$ is determined
 in the same way as in the anti-tagged case. 
 It is found that the influence of the charm quark mass and
 fragmentation function is small,
 and the relative uncertainty on ${\cal R}_{\rm tag}^{\rm MC}$ and thus 
 on the averaged cross-section $\sigcc$ is only 5$\%$.

 In contrast, for $x<0.1$, the Monte Carlo generators predict very
 different extrapolation factors due to the large discrepancy between 
 the predicted invisible part of the cross-section.
 For $x<0.1$, the HERWIG extrapolation factor,
 ${\cal R}_{\rm tag}^{\rm MC}=12.9$, is more than twice as large as
 the Vermaseren factor ${\cal R}_{\rm tag}^{\rm MC}=5.1$.   
 The predicted cross-sections of the Monte Carlo models and
 the NLO calculation of Laenen et al.~\cite{bib-Laenen}
 are given in Table~\ref{tab-tagmc}.
 Since the hadron-like contribution is neglected in the QPM,
 the Vermaseren cross-section is much smaller than the 
 LO and the NLO cross-section for $x<0.1$.
 In contrast, mainly due to the massless approach taken, the 
 prediction from HERWIG is higher than the cross-section
 from the LO and the NLO calculation.
 Therefore it is likely that the correct cross-section,
 and therefore the correct extrapolation factor, lies within 
 the range of the two Monte Carlo predictions.
 Half the difference between the two extrapolated cross-sections is taken
 into account as extrapolation error on the averaged cross-section $\sigcc$.

%
 Finally, the value of the charm structure function $\XQCF2$ of the photon, 
 averaged
 over the corresponding bin in $x$, is determined by 
\begin{equation}
\XQCF2=\sigcc\cdot\left(\frac{\XQCF2}{\sigcc}\right)_{\rm NLO},
\label{eq-nlo}
\end{equation}
 where the ratio $(\XQCF2/\sigcc)_{\rm NLO}$ is given
 by the NLO calculation of Laenen et al.~\cite{bib-Laenen}.
 The mean virtuality in the measured region 
 $5$~GeV$^2<Q^2<100$~GeV$^2$ is about
 $\langle Q^2\rangle\approx 20~{\rm GeV}^2$, in agreement with
 the values from the generated HERWIG and Vermaseren Monte Carlo events.
 The $\XQCF2$ values are given in
 Table~\ref{tab-tagda}. They are calculated from the 
 individual charm cross-sections obtained using the HERWIG and
 Vermaseren models and from the averaged cross-section.
 \par
\begin{figure}[htbp]
\vspace*{-2.cm}
   \begin{center}
      \mbox{
          \epsfxsize=18.5cm
          \epsffile{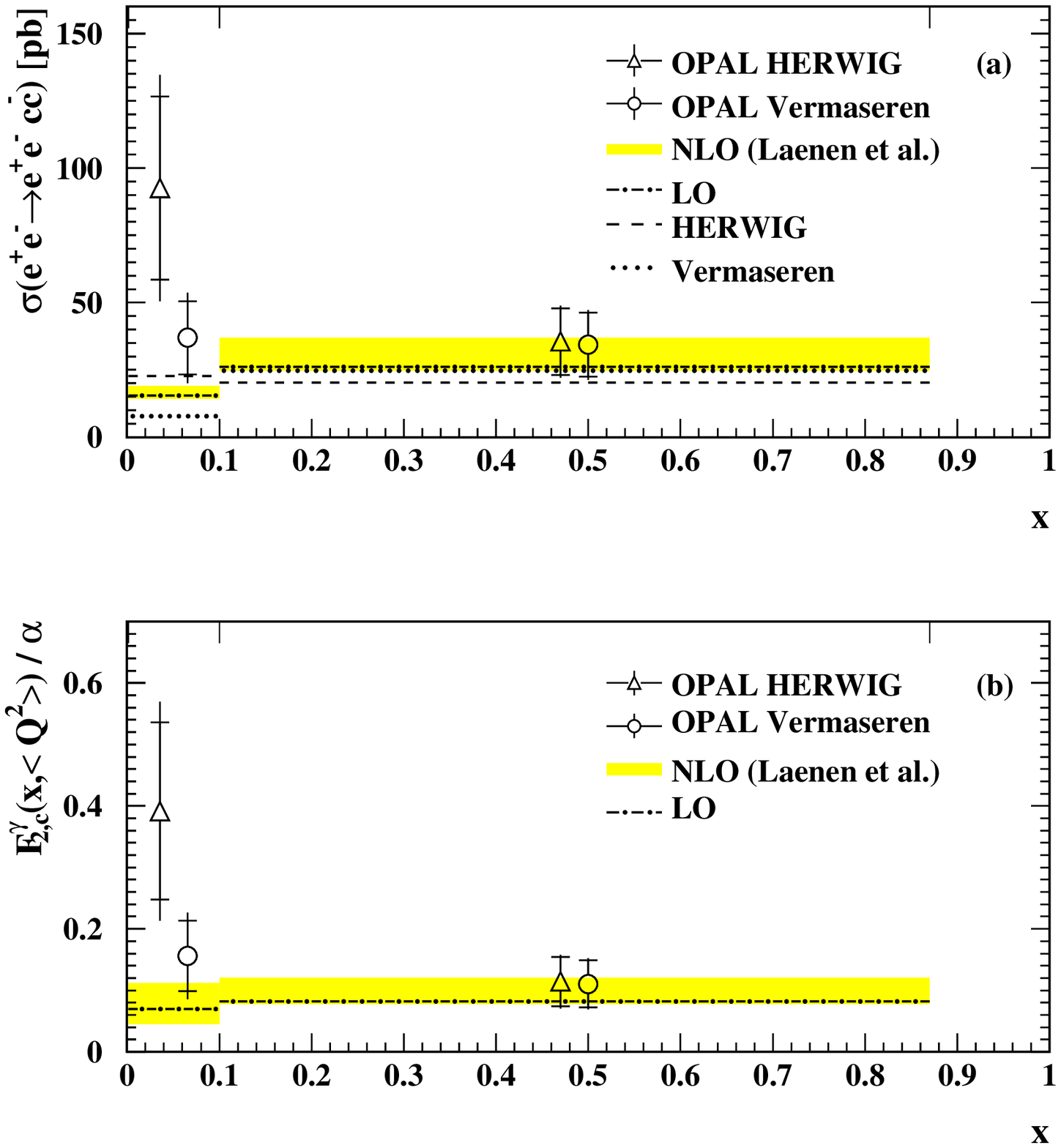}
           }
   \end{center}
\caption{Results compared with predictions for a) the deep inelastic 
         electron-photon scattering cross-section
         $\sigcc$, with $5~{\rm GeV}^2<Q^2<100~{\rm GeV}^2$
         and b) for the charm structure function of the photon divided
         by the fine structure constant, $\XQCF2/\alpha$ at
         $\langle Q^2\rangle=20~{\rm GeV}^2$.
         The data are shown individually corrected with the HERWIG
         and Vermaseren Monte Carlo generators. 
         The inner error bar is the statistical and 
         the outer error bar is the full error.
         The measurements are presented at the central $x$ values of the bins.
         The results obtained with the HERWIG and Vermaseren generators
         are slightly separated for a better visibility.  
         The calculation of Laenen et al.~\cite{bib-Laenen} is performed in
         LO and NLO. The band for the NLO calculation
         indicates the theoretical uncertainties assessed by varying the
         charm quark mass and renormalisation and factorisation scales.
         }
\label{fig-xmvind}
\end{figure}

 In Fig.~\ref{fig-xmvind}~a), the measured cross-sections
 obtained using the individual Monte Carlo
 models, are compared to the calculation of Laenen et al.~\cite{bib-Laenen}
 performed in LO and NLO and to the Monte Carlo results, 
 and Fig.~\ref{fig-xmvind}~b) shows the charm structure function $\CF2$.
 The NLO prediction is based on $m_{\rm c}=1.5~{\rm GeV}$ and
 the renormalisation and factorisation scales are chosen to
 be $\mu_{\rm R} = \mu_{\rm F} = Q$. 
 The calculation is obtained for the sum of the point-like and
 hadron-like contributions to $\CF2$, using the GRV-NLO parametrisation
 in the calculation of the hadron-like part.
 The NLO corrections are predicted to be small for the whole $x$ range.
 The NLO calculation is shown as a band representing the uncertainty of 
 the theoretical prediction, evaluated by varying the charm quark mass
 between 1.3 and 1.7~GeV and by changing the renormalisation and
 factorisation scales in the range $Q/2 \le \mu_{\rm R}=\mu_{\rm F}\le 2~Q$.
 \par
 For $x>0.1$, all cross-section predictions in Fig.~\ref{fig-xmvind} a)
 are consistent with one another.
 Because the cross-section prediction from the Vermaseren model is 
 consistent with the point-like contribution to the LO calculation for the
 whole $x$ range, the contributions from longitudinal photons are expected
 to be small.
 For $x<0.1$, the situation is different. The NLO calculation predicts
 the hadron-like and point-like component to be of about equal size.
 Therefore the purely point-like QPM prediction of the Vermaseren model is
 expected to underestimate the data if a hadron-like contribution
 exists.
 The HERWIG Monte Carlo predicts the highest cross-section, which 
 is expected, since the massless approach should overestimate the 
 cross-section, as explained in Section~4.
\par
 The different behaviour of the Monte Carlo cross-sections in the two
 regions of $x$ is reflected in the measured cross-sections
 shown in Fig.~\ref{fig-xmvind}~a).
 For $x>0.1$ the individual measured cross-sections obtained by correcting
 with HERWIG and Vermaseren are very similar,
 the error of the measured cross-section is dominated by the statistical
 uncertainty, and the NLO calculation is in good agreement with the data.
 In contrast, for $x<0.1$, the result suffers from the strong model 
 dependence discussed above. The result based on the HERWIG 
 generator is much higher than the result obtained using the Vermaseren
 model.
 Despite this uncertainty the corrected data suggest a 
 cross-section which is above the purely point-like component, i.e.
 the hadron-like component of $\CF2$ is non-zero.
 This observation is independent of the Monte Carlo model chosen for
 correction.
 Averaging the individual results is therefore safe for 
 $x>0.1$, but for $x<0.1$ the averaged result suffers from 
 large model uncertainties and has to be interpreted with care.
 In Fig.~\ref{fig-xmvind}~b) the cross-section measurements are converted 
 into the measured charm structure function using Eq.~\ref{eq-nlo}.
 The conclusions derived from $\CF2$ and from the cross-sections are the same. 

\begin{figure}[htbp]
\vspace*{-3.cm}
   \begin{center}
      \mbox{
          \epsfxsize=18.5cm
          \epsffile{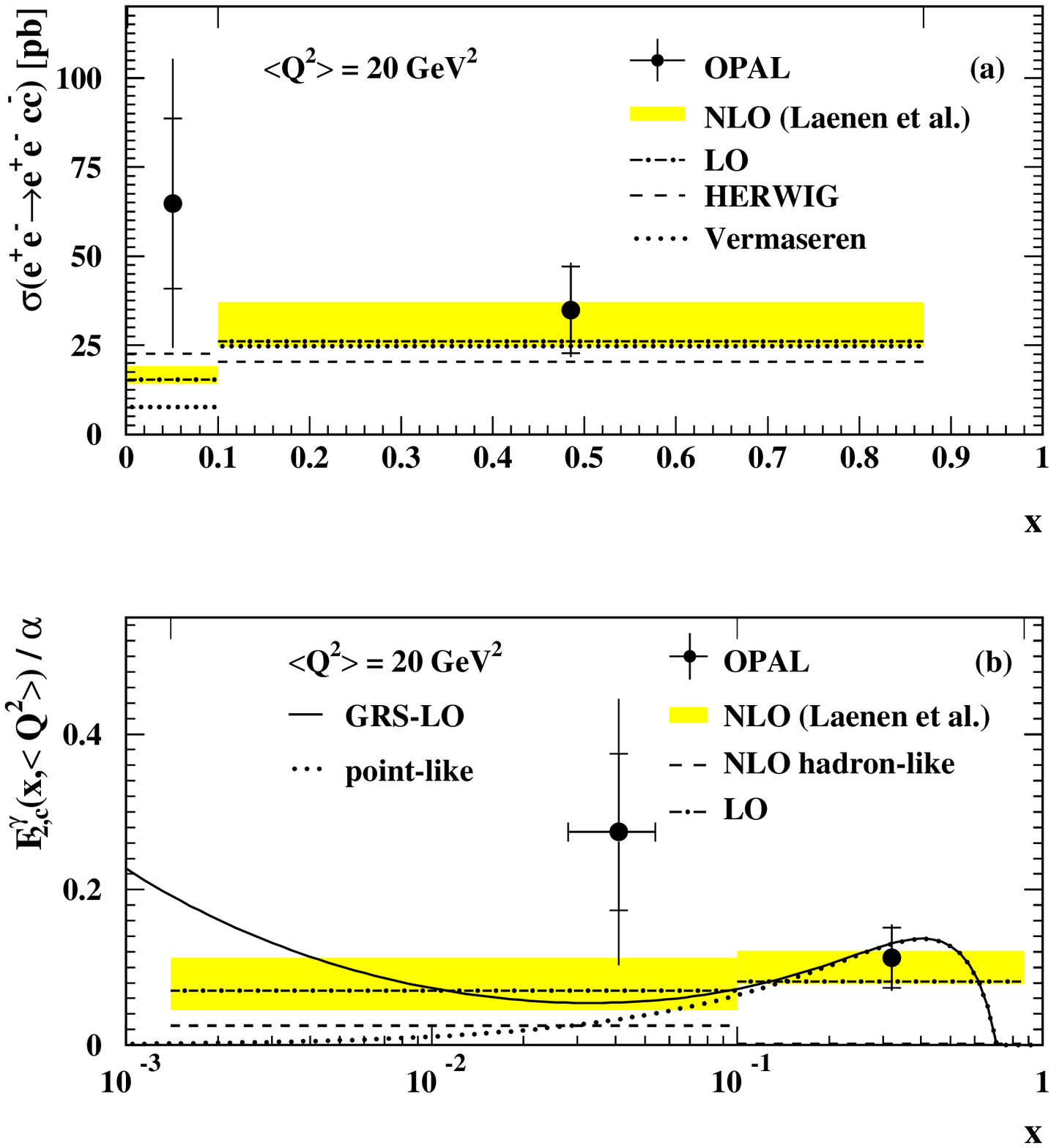}
           }
   \end{center}
\caption{OPAL results for a) the deep inelastic electron-photon scattering
         cross-section $\sigcc$, with $5~{\rm GeV}^2<Q^2<100~{\rm GeV}^2$ and
         b) for the charm structure function of the photon divided
         by the fine structure constant, $\XQCF2/\alpha$, for
         an average $\langle Q^2\rangle$ of $20~{\rm GeV}^2$.
         The data points are obtained averaging the results obtained
         with the HERWIG and Vermaseren Monte Carlo models.
         The outer error bar is the total error and the inner error
         bar the statistical error.
         The $x$ values of the data points are obtained by averaging the
         mean $x$ values taken from the HERWIG and Vermaseren generators.  
         The data are compared to the calculation of
         Laenen et al.~\cite{bib-Laenen} performed in LO and NLO. 
         The band for the NLO calculation
         indicates the theoretical uncertainties assessed by varying the
         charm quark mass and renormalisation and factorisation scales. 
         In a) the cross-section predictions of the Monte Carlo
         generators HERWIG and Vermaseren are also given.
         b) also shows the prediction of the GRS-LO parametrisation 
         for the whole structure function at $\langle Q^2\rangle
=20~{\rm GeV}^2$ 
         and the point-like component alone.}
\label{fig-xresult}
\end{figure}

 In Fig.~\ref{fig-xresult}, the averaged results are presented.
 Fig.~\ref{fig-xresult}~a) shows the cross-section on a linear scale in 
 $x$ in comparison to the same predictions as in Fig.~\ref{fig-xmvind}.
 In Fig.~\ref{fig-xresult}~b) the charm structure function is presented
 on a logarithmic scale in $x$ for $\langle Q^2\rangle=20~{\rm GeV}^2$.
 The data points for $\CF2$ are located at the mean value of $x$,
 denoted with $\langle x\rangle$.
 The values are the averaged $\langle x\rangle$ values obtained with both
 Monte Carlo generators, and half the difference of the HERWIG and
 Vermaseren predictions is taken as the uncertainty.
 For $x>0.1$, the predicted $\langle x\rangle$ is around 0.32 and the
 difference between the HERWIG and Vermaseren programs is invisible.  
 The point-like contribution decreases for decreasing $x$, whereas the
 hadron-like component rises.
 Consequently, for $x<0.1$, the HERWIG Monte Carlo predicts a smaller 
 average value of $\langle x\rangle=0.028$ than the Vermaseren Monte Carlo
 which yields $\langle x\rangle=0.054$.
 \par 
 In addition to the full NLO prediction, the predicted hadron-like component of
 $\CF2$ is also shown in Fig.~\ref{fig-xresult}~b).
 This contribution is very small for $x>0.1$ and therefore in this 
 range the NLO calculation is an almost purely perturbative prediction with
 the charm quark mass and the strong coupling constant as the only
 free parameters. This prediction nicely describes the data.
 To illustrate the shape of $\CF2$ the data are also compared to the
 GRS-LO~\cite{bib-grs} prediction and to the point-like component alone
 both shown for $Q^2=20$~GeV$^2$.
 The point-like component strongly decreases for decreasing $x$.
 The full $\CF2$ evaluated at $Q^2=20$~GeV$^2$ agrees with the data.
 The change of $\CF2$ within the range of $Q^2$ studied is large.
 The maximum value of $\CF2$ for $x>0.1$ rises by about
 a factor of five between $Q^2=5$~GeV$^2$ and $Q^2=100$~GeV$^2$ 
 and the charm thresholds moves from about $x=0.35$ to about $x=0.9$.
 \par
 In conclusion, for $x>0.1$, the purely perturbative NLO calculation is
 in good agreement with the measurement and for $x<0.1$, 
 the measurement suffers from large uncertainties of
 the invisible cross-section predicted by the HERWIG and Vermaseren
 Monte Carlo models, and therefore the result is not very precise.
 However, despite the large error in this region the data suggest a 
 non-zero hadron-like component of $\CF2$.
%
\section{Conclusion}
%
 We have measured the inclusive production of $\DSTPM$ mesons in 
 photon-photon collisions using the OPAL detector at LEP at 
 \epem~centre-of-mass energies $\sqee=183$ and $189$~GeV.
 The $\DSTP$ mesons are reconstructed in their decay to
 ${\rm D}^0\pi^+$ with the ${\rm D}^0$ observed in the two
 decay modes ${\rm K}^-\pi^+$ and ${\rm K}^-\pi^+\pi^-\pi^+$.
 In total, $100.4\pm12.6~(\rm stat)~\DST$ mesons are selected in anti-tagged
 events and $\ntag~\DST$ mesons in single-tagged events.
 \par
 In the anti-tagged event sample, the direct and single-resolved contributions
 are separated using di-jet events reconstructed with a cone
 jet finding algorithm, and for all observed events by fitting the
 distribution of the scaled $\DST$ transverse momentum $\xt$.
 Both methods yield consistent results, and due to the larger statistics used,
 the second method is more precise.
 It is found that
 in the kinematical region $\ptdst>2~{\rm GeV}$ and $|\etadst|<1.5$
 the direct contribution to the process \mbox{\epem $\to$ \epem $\DST X$} is 
 $(51\pm10)\%$ and that the single-resolved contribution is $(49\pm10)\%$. 
 \par
 Differential cross-sections as functions of the $\DST$ transverse momentum
 and pseudorapidity are measured for anti-tagged events and are compared to 
 a NLO calculation by  Kniehl et al.~\cite{bib-Kniehl} using the massless 
 approach, and by Frixione et al.~\cite{bib-Frixione} using the massive
 approach. 
 It is found that despite the low values of $\ptdst$ studied 
 the massless calculation is in good agreement with the data.
 The massive calculation agrees with the measured cross-section for
 $\ptdst>3~{\rm GeV}$ but underestimates the data for lower values of
 $\ptdst$.
 \par
 The total cross-section of the process \epem $\to$ \epem $\ccbar$, where
 the charm quarks are produced in the collision of two quasi-real photons,
 is measured to be $\sigcc = \sigtot$, with a direct contribution of
 $\sigcc_{\rm dir} = \sigdir$ and a single-resolved 
 contribution of $\sigcc_{\rm res} = \sigres$. 
 The NLO calculation of Ref.~\cite{bib-DKZZ} and the measurements by
 L3~\cite{bib-L3} are in agreement with this result.
 \par
 The first measurement of the charm structure function $\XQCF2$ of the photon
 has been performed based on $\ntag~\DST$ mesons 
 reconstructed in single-tagged events. The value of $\XQCF2$ is determined
 for an average $\langle Q^2\rangle$ of $20~{\rm GeV}^2$ and
 in two regions of $x$, $0.0014<x<0.1$ and $0.1<x<0.87$.
 The NLO corrections to $\XQCF2$ are predicted to be small for all
 $x$ and the contribution of the hadron-like component is negligible
 for $x>0.1$, which means that $\XQCF2$ can be predicted purely 
 perturbatively in this region.
 For $x>0.1$, the perturbative NLO calculation of Laenen et
 al.~\cite{bib-Laenen} is in good agreement with the measurement.
 For $x<0.1$, the measurement suffers from large uncertainties of
 the invisible cross-section predicted by the HERWIG and Vermaseren
 Monte Carlo models, and therefore the result is not very precise.
 However, despite the large error in this region the data suggest a 
 non-zero hadron-like component of $\CF2$.

\medskip
\bigskip\bigskip\bigskip
\appendix
\par
\section*{Acknowledgements}
\par
 We especially wish to thank Michael Kr\"amer and Eric Laenen
 for many interesting and valuable discussions and good collaboration. 
 We are grateful to them, to Stefano Frixione and to Bernd Kniehl
 for providing their NLO calculations.\\
We particularly wish to thank the SL Division for the efficient operation
of the LEP accelerator at all energies
 and for their continuing close cooperation with
our experimental group.  We thank our colleagues from CEA, DAPNIA/SPP,
CE-Saclay for their efforts over the years on the time-of-flight and trigger
systems which we continue to use.  In addition to the support staff at our own
institutions we are pleased to acknowledge the  \\
Department of Energy, USA, \\
National Science Foundation, USA, \\
Particle Physics and Astronomy Research Council, UK, \\
Natural Sciences and Engineering Research Council, Canada, \\
Israel Science Foundation, administered by the Israel
Academy of Science and Humanities, \\
Minerva Gesellschaft, \\
Benoziyo Center for High Energy Physics,\\
Japanese Ministry of Education, Science and Culture (the
Monbusho) and a grant under the Monbusho International
Science Research Program,\\
Japanese Society for the Promotion of Science (JSPS),\\
German Israeli Bi-national Science Foundation (GIF), \\
Bundesministerium f\"ur Bildung, Wissenschaft,
Forschung und Technologie, Germany, \\
National Research Council of Canada, \\
Research Corporation, USA,\\
Hungarian Foundation for Scientific Research, OTKA T-029328, 
T023793 and OTKA F-023259.\\

\newpage

\vspace*{2.cm}         
\begin{table}[htpb]
\begin{center}

\begin{tabular}{|c|c|c|}
\hline
\rule[-.25cm]{0.cm}{.8cm} & $\D0\to \K^- \pi^+$ & $\D0\to \K^-\pi^+\pi^-\pi^+$  \\ \hline\hline
\rule[-.25cm]{0.cm}{.8cm} direct  &  $(40.8\pm1.7)\%$ & $(14.4\pm0.8)\%$  \\ \hline
\rule[-.25cm]{0.cm}{.8cm}single-resolved & $(38.1\pm2.0)\%$ & $(11.2\pm0.9)\%$  \\ \hline
\end{tabular}
\caption{$\DST$ selection efficiencies $\epsilon$ for the two decay modes 
         and for direct and single-resolved events as 
         determined from the PYTHIA Monte Carlo (for anti-tagged
         events only). 
         The selection efficiencies refer to $\DST$
         mesons with $\ptdst>2~{\rm GeV}$  and $|\etadst| < 1.5$.
         Only statistical errors are given.}
\label{tab-eff}
\vspace*{3.cm}

\hspace*{-1.cm}
\begin{tabular}{|c|c|c|c|c|c|c|}
\hline
\rule[-.25cm]{0.cm}{1.cm} $\ptdst$ & $\langle\ptdst\rangle$ & \multicolumn{2}{c|}{$\ndst$} & \multicolumn{3}{c|}{$\dspt$~[pb/GeV]} \\ \cline{3-7} 
\rule[-.25cm]{0.cm}{.8cm}$[{\rm GeV}]$ & $[{\rm GeV}]$  & $ \K^- \pi^+$ & $ \K^-\pi^+\pi^-\pi^+$ & $ \K^- \pi^+$ & $ \K^-\pi^+\pi^-\pi^+$ & combined\\  \hline \hline
\rule[-.25cm]{0.cm}{.8cm} $2-3$  & $2.46$ & $42.9\pm7.7$ & $27.4\pm7.3$ & $20.6\pm3.7~\pm3.0$ & $21.5\pm5.8~\pm~6.1$ & $20.8\pm3.1\pm~2.4$ \\ \hline
\rule[-.25cm]{0.cm}{.8cm} $3-5$  & $3.82$ & $18.4\pm4.5$ & $11.4\pm4.2$ & $~3.7\pm0.9~\pm0.5$ & $~3.1\pm1.2~\pm~0.7$ & $~3.5\pm0.7\pm~0.3$\\ \hline
\rule[-.25cm]{0.cm}{.8cm} $5-12$ & $7.30$ & $~8.3\pm3.0$ & $~4.5\pm2.2$ & $0.38\pm0.14\pm0.06$ & $0.25\pm0.12\pm0.05$ & $0.31\pm0.09\pm0.09$\\ \hline
\end{tabular}
\caption {Number of reconstructed $\DST$ mesons with $|\etadst|<1.5$ 
          in bins of $\ptdst$ for both decay modes after
          background subtraction (for anti-tagged events only).
          The differential $\DST$ cross-section as a function 
          of $\ptdst$ for each decay mode and the combined cross-section is
          also given.
          The first error is statistical and the second error is systematic.}
\label{tab-sigpt}

\end{center}
\end{table}

\newpage
\begin{table}[htpb]
\begin{center}
\vspace*{-1.cm}
\begin{tabular}{|c|c|c|c|c|c|}
\hline
\rule[-.25cm]{0.cm}{.8cm} \raisebox{-2.ex}{$|\etadst|$} & \multicolumn{2}{c|}{$\ndst$} & \multicolumn{3}
{c|}{$\dseta$~[pb]} \\ \cline{2-6}
\rule[-.25cm]{0.cm}{.8cm} & $ \K^- \pi^+$ & $ \K^-\pi^+\pi^-\pi^+$ & $ \K^- \pi^+$ & $ \K^-\pi^+\pi^-\pi^+$ & combined  \\  \hline \hline
\rule[-.25cm]{0.cm}{.8cm} $0.0-0.5$ & $29.1\pm5.9$ & $17.0\pm5.3$ & $22.6\pm4.6\pm2.7$ & $18.0\pm~5.6\pm4.7$ & $21.0\pm3.5\pm2.1$ \\ \hline
\rule[-.25cm]{0.cm}{.8cm} $0.5-1.0$ & $18.1\pm5.3$ & $24.0\pm5.8$ & $14.5\pm4.2\pm3.8$ & $23.8\pm~5.8\pm4.9$ & $18.0\pm3.4\pm2.8$ \\ \hline
\rule[-.25cm]{0.cm}{.8cm} $1.0-1.5$ & $22.8\pm5.0$ & $~6.3\pm3.7$ & $25.3\pm5.6\pm2.9$ & $19.5\pm11.7\pm6.1$ & $24.2\pm5.0\pm2.2$ \\ \hline
\end{tabular}
\caption {Number of reconstructed $\DST$ mesons with 
          $2~{\rm GeV}<\ptdst<12~{\rm GeV}$ in bins of $|\etadst|$ for both
          decay modes after background subtraction (for anti-tagged events
          only).
          The differential $\DST$ cross-section as a function
          of $|\etadst|$ for each decay mode and the combined cross-section 
          is also given. 
          The first error is statistical and the second error is systematic.}
\label{tab-sigeta}
\vspace*{1.cm}


\begin{tabular}{|c|c|c|c|c|} \hline
\rule[-.45cm]{0.cm}{1.2cm}       & $\sigma^{\DST}_{\rm dir}$~[pb] & \multicolumn{3}{c|}{$\sigma^{\DST }_{\rm res}$~[pb]} \\ \hline \hline
\rule[-.25cm]{0.cm}{.8cm}       &        & ~~3.8 - 5.5~~ & ~6.9 - 9.9~ & 
23.0 - 37.0  \\ \
\rule[-.25cm]{0.cm}{.8cm}\raisebox{2.ex}[-2.ex]{PYTHIA} & 
\raisebox{2.ex}[-2.ex]{14.0 - 14.9}  & (SaS-1D)   & (GRV)    &  (LAC1)  \\ \hline 
\end{tabular}
\caption{
Predicted integrated LO cross-section of the
process \epem $\to$ \epem $\DST X$ in the kinematical region
$2~{\rm GeV}<\ptdst<12~{\rm GeV}$ and $|\etadst|<1.5$
calculated with the PYTHIA Monte Carlo using
different parametrisations of the parton densities (anti-tagged events only).
Direct and single-resolved cross-sections are given separately.
The charm quark mass was varied between 1.3 and 1.7 GeV.}
\label{tab-sigma}

\vspace*{1.cm}
\begin{tabular}{|c|c|c|c|c|} \hline
\rule[-.25cm]{0.cm}{.8cm} \raisebox{-2.ex}{$x$} & \raisebox{-2.ex}{$N_{\DST}^{\rm rec}$} & \multicolumn{2}{c|}{$\sigma_{\rm tag}^{\DST}$~[pb], corrected with} & {$\sigma_{\rm tag}^{\rm \DST}$~[pb]} \\ \cline{3-5}
\rule[-.25cm]{0.cm}{.8cm}     &                      & HERWIG   & Vermaseren                    &   average          \\ \hline \hline
\rule[-.25cm]{0.cm}{.8cm} $0.0014-0.1$ & $\pz9.9\pm3.6$ & $3.1\pm1.1\pm0.7$ & $4.7\pm1.8\pm0.9$     & $3.9\pm1.4\pm1.0$ \\ \hline
\rule[-.25cm]{0.cm}{.8cm} $0.1-0.87$ & $20.0\pm4.7$ & $4.0\pm1.4\pm0.7$ & 
$4.0\pm1.4\pm0.7$     & $4.0\pm1.4\pm0.6$    \\ \hline 
\end{tabular}
\caption{Number of reconstructed $\DST$ mesons, $N_{\DST}^{\rm rec}$, 
         found in tagged events and  
         $\sigma_{\rm tag}^{\DST}$, obtained by correcting with HERWIG 
         and Vermaseren in two bins of $x$, where
         $\sigma_{\rm tag}^{\DST}$ is the deep inelastic
         electron-photon scattering cross-section for $\DST$ production in 
         the restricted kinematic range as defined in the text.
         The first error on $\sigma_{\rm tag}^{\DST} $ is the
         statistical error of the data and the second error is
         the systematic error.
         The right hand column gives the averaged 
         cross-sections obtained by correcting with the two Monte Carlo models.}
\label{tab-unfold}
\end{center}
\end{table}

\newpage
\begin{table}[htpb]
\begin{center}
\begin{tabular}{|c||c|c||c|} \hline
\rule[-.25cm]{0.cm}{1.cm} \raisebox{-2.ex}{$x$} & \multicolumn{2}{c||}{$\sigcc$ [pb], corrected with} & $\sigcc$ [pb] \\ \cline{2-4}
\rule[-.25cm]{0.cm}{.8cm} & HERWIG & Vermaseren & average  \\ \hline \hline
\rule[-.25cm]{0.cm}{.8cm} $0.0014-0.1$ & $92.6\pm34.1\pm24.6$ & $36.9\pm13.6\pm9.8$ & $64.8\pm23.9\pm17.2\pm27.9$ \\ \hline 
\rule[-.25cm]{0.cm}{.8cm} $0.1-0.87$   & $35.5\pm12.4\pm~5.2$ & $34.3\pm11.9\pm5.1$ & $34.9\pm12.1\pm~5.1\pm\pz1.7$ \\  \hline \hline
\rule[-.25cm]{0.cm}{1.cm} \raisebox{-2.ex}{$x$} & \multicolumn{2}{c||}{$\XQCF2/\alpha$, corrected with} & $\XQCF2/\alpha$ \\ \cline{2-4}
\rule[-.25cm]{0.cm}{.8cm} & HERWIG & Vermaseren & average  \\ \hline \hline
\rule[-.25cm]{0.cm}{.8cm} $0.0014-0.1$ & $0.39\pm0.14\pm0.10$ & $0.16\pm0.06\pm0.04$ & $0.27\pm0.10\pm0.07\pm0.12$ \\ \hline
\rule[-.25cm]{0.cm}{.8cm} $0.1-0.87$   & $0.11\pm0.04\pm0.02$ & $0.11\pm0.04\pm0.02$ & $0.11\pm0.04\pm0.02\pm0.01$ \\ \hline
\end{tabular}
\caption{Total charm cross-section in deep inelastic electron-photon
         scattering, $\sigcc$, for $5~{\rm GeV}^2<Q^2<100~{\rm GeV}^2$
         and the charm structure function of the photon divided by the fine
         structure constant, $\XQCF2/\alpha$, averaged over the corresponding
         bin in $x$ for $\langle Q^2\rangle=20~{\rm GeV}^2$. 
         The cross-section and $\XQCF2/\alpha$ 
         are presented corrected using both the 
         HERWIG and Vermaseren Monte Carlo models.
         The averaged values for the cross-section and $\XQCF2/\alpha$ are
         also given.
         The first errors are 
         statistical, the second errors systematic, and
         the third errors are the extrapolation uncertainties.
         }
\label{tab-tagda}

\vspace{2.cm}
\begin{tabular}{|c||c|c|c|c||c|c|} \hline
\rule[-.25cm]{0.cm}{1.cm} \raisebox{-2.ex}{$x$} & \multicolumn{4}{c||}{$\sigcc$ [pb]} & \multicolumn{2}{c|}{$\XQCF2/\alpha$} \\ \cline{2-7}
\rule[-.25cm]{0.cm}{.8cm} & HERWIG & Vermaseren & LO & NLO & LO & NLO  \\ \hline \hline
\rule[-.25cm]{0.cm}{.8cm} $0.0014-0.1$ & 22.6 & ~7.7 & 15.3 & $16.3^{+2.8}_{-2.1}$ & 0.070 & $0.069^{+0.043}_{-0.024}$\\ \hline
\rule[-.25cm]{0.cm}{.8cm} $0.1-0.87$   & 20.3 & 24.7 & 26.1 & $30.1^{+6.9}_{-5.5}$ & 0.082 & $0.097^{+0.024}_{-0.019}$\\ \hline
\end{tabular}
\caption{Predicted total charm cross-section in deep inelastic
         electron-photon scattering, $\sigcc$, for 
         $5~{\rm GeV}^2<Q^2<100~{\rm GeV}^2$ according to the
         Monte Carlo generators HERWIG and Vermaseren and according
         to the calculation of Laenen et al.~\cite{bib-Laenen} performed in
         LO and NLO. For the calculation, also the $\XQCF2/\alpha$
         prediction is quoted in LO and NLO for $Q^2=20~{\rm GeV}^2$.
         The errors of the NLO results are obtained by varying the charm
         quark mass and the renormalisation and factorisation scales
         in the calculation.} 
\label{tab-tagmc}
\end{center}
\end{table}

\newpage

\end{document}